\documentclass[final,1p,times]{elsarticle}
\usepackage{amssymb}
\usepackage{amsfonts}
\usepackage{amsmath}
\usepackage{geometry}
\usepackage{bm}
\numberwithin{equation}{section}
\newtheorem{thm}{Theorem}
\newtheorem{lem}[thm]{Lemma}
\newdefinition{rmk}{Remark}
\newtheorem{defn}{Definition}
\newproof{pf}{Proof}
\newproof{pot}{Proof of Theorem  
}
\begin{document}

\begin{frontmatter}
\title{The generating functions of Lame equation in  Weierstrass's form}

\author{Yoon Seok Choun\corref{cor1}}
\ead{Yoon.Choun@baruch.cuny.edu; ychoun@gc.cuny.edu; ychoun@gmail.com}
\cortext[cor1]{Correspondence to: Baruch College, The City University of New York, Natural Science Department, A506, 17 Lexington Avenue, New York, NY 10010} 
\address{Baruch College, The City University of New York, Natural Science Department, A506, 17 Lexington Avenue, New York, NY 10010}
\begin{abstract}

Lame equation arises from deriving Laplace equation in ellipsoidal coordinates; in other words, it's called ellipsoidal harmonic equation.  Lame function is applicable to diverse areas such as boundary value problems in ellipsoidal geometry, chaotic Hamiltonian systems, the theory of Bose-Einstein condensates, etc.

By applying generating function into modern physics (quantum mechanics, thermodynamics, black hole, supersymmetry, special functions, etc), we are able to obtain the recursion relation, a normalization constant for the wave function and expectation values of any physical quantities. For the case of hydrogen-like atoms, generating function of associated Laguerre polynomial has been used in order to derive expectation values of position and momentum. By applying integral forms of Lame polynomial in   Weierstrass's form in which makes $B_n$ term terminated\cite{Chou2012g}, I consider generating function of it including all higher terms of $A_n$'s.\footnote{`` higher terms of $A_n$'s'' means at least two terms of $A_n$'s.}

This paper is 8th out of 10 in series ``Special functions and three term recurrence formula (3TRF)''. See section 4 for all the papers in the series.  Previous paper in series deals with the power series expansion and the integral formalism of Lame equation in   Weierstrass's form and its asymptotic behavior \cite{Chou2012g}. The next paper in the series describes analytic solution for grand confluent hypergeometric function\cite{Chou2012i}.
\end{abstract}

                                      
\end{frontmatter}  
 
\section{Introduction}
In 1837, Gabriel Lame introduced second ordinary differential equation which has four regular singular points in the method of separation of variables applied to the Laplace equation in elliptic coordinates\cite{Lame1837}. Various authors has called this equation as `Lame equation' or `ellipsoidal harmonic equation'\cite{Erde1955}. 
Lame function is applicable to diverse areas such as boundary value problems in ellipsoidal geometry, chaotic Hamiltonian systems, the theory of Bose-Einstein condensates, etc.

Lame equation is the special case of Heun's equation.
In Ref.\cite{Heun1889}, Heun's equation is a second-order linear ordinary differential equation of the form
\begin{equation}
\frac{d^2{y}}{d{x}^2} + \left(\frac{\gamma }{x} +\frac{\delta }{x-1} + \frac{\epsilon }{x-a}\right) \frac{d{y}}{d{x}} +  \frac{\alpha \beta x-q}{x(x-1)(x-a)} y = 0 \label{eq:50}
\end{equation}
As we compare (\ref{eq:2}) with (\ref{eq:50}), all coefficients on the above are correspondent to the following way.
\begin{equation}
\begin{split}
& \gamma ,\delta ,\epsilon  \longleftrightarrow   \frac{1}{2} \\ & a\longleftrightarrow  \rho ^{-2} \\ & \alpha  \longleftrightarrow \frac{1}{2}(\alpha +1) \\
& \beta   \longleftrightarrow -\frac{1}{2} \alpha \\
& q \longleftrightarrow  -\frac{1}{4}h \rho ^{-2} \\ & x \longleftrightarrow \xi = sn^2(z,\rho ) 
\end{split}\label{eq:51}   
\end{equation}
Associated Laguerre function, the special case of Hypergeometric function, arises from deriving the Laplace equation in the spherical coordinates system generally.
Lame and Heun functions arise from deriving the Laplace equation in general Jacobi ellipsoidal or conical coordinates.\cite{Kalnin2005}  Also Heun polynomial arises in the separable coordinate systems on the n-sphere.\cite{Kaln1990} 

According to E. G. Kalnins and W. Miller Jr.(1990 \cite{Kaln1990}) , ``Lame and Heun functions have received relatively little attention, since they are rather intractable. Unfortunately the beautiful identities appearing have received little notice, probably because the methods of proof seemed obscure.'' 
A. Erdelyi also mentioned, ``The connections between these integral equations and the question has not yet been dealt with whether the known integral equations exhaust all possible types of integral equations connected with Lame polynomials.  In fact it is very plausible that they do not.''(Erdelyi 1940)\cite{Erde1940} 

In Ref.\cite{Chou2012g}, I show power series expansion in closed forms of Lame function in   Weierstrass's form and its representation of the form of integrals for the cases of infinite series and polynomial in which makes $B_n$ term terminated. I show that a $_2F_1$ function recurs in each of sub-integral forms of Lame function in   Weierstrass's form: the first sub-integral form contains zero term of $A_n's$, the second one contains one term of $A_n$'s, the third one contains two terms of $A_n$'s, etc. And I show asymptotic expansions of Lame function for infinite series and the special case as $\rho \approx 0$.  

In this paper, I consider the generating function of Lame polynomial in   Weierstrass's form in which makes $B_n$ term terminated. Since the generating function of Lame polynomial is derived, we might be possible to construct orthogonal relations of Lame polynomial.
In the physical point of view we might be possible to obtain the normalized constant for the wave function in modern physics, recursion relation and its expectation value of any physical quantities from the generating function of Lame polynomial. For the case of hydrogen-like atoms, the normalized wave function is derived from the generating function of associated Laguerre polynomial. And the expectation value of physical quantities such as position and momentum is constructed by applying the recursive relation of associated Laguerre polynomial.

There are three types of polynomials in three-term recurrence relation of a linear ordinary differential equation: (1) polynomial which makes $B_n$ term terminated: $A_n$ term is not terminated, (2) polynomial which makes $A_n$ term terminated: $B_n$ term is not terminated, (3) polynomial which makes $A_n$ and $B_n$ terms terminated at the same time.\footnote{If $A_n$ and $B_n$ terms are not terminated, it turns to be infinite series.} In general Lame polynomial (or Lame spectral polynomial) is defined as type 3 polynomial where $A_n$ and $B_n$ terms terminated. Lame polynomial comes from a Lame equation that has a fixed integer value of $\alpha $, just as it has a fixed value of $h$. In three-term recurrence formula, polynomial of type 3 I categorize as complete polynomial. In future papers I will derive type 3 Lame polynomial. In this paper I construct the generating function for Lame polynomial of type 1: I treat the spectral parameter $h$ as a free variable and $\alpha $ as a fixed value. In my next papers I will work on the generating functions for Lame polynomial of type 2.

The Lame equation in Weierstrass's form is
\begin{equation}
\frac{d^2{y}}{d{z}^2} = \{ \alpha (\alpha +1)\rho^2\;sn^2(z,\rho )-h\} y(z)\label{eq:1}
\end{equation}
where $\rho$, $\alpha $  and $h$ are real parameters such that $0<\rho <1$ and $\alpha \geq -\frac{1}{2}$.
If we take $sn^2(z,\rho )=\xi $ as independent variable, Lame equation becomes
\begin{equation}
\frac{d^2{y}}{d{\xi }^2} + \frac{1}{2}\left(\frac{1}{\xi } +\frac{1}{\xi -1} + \frac{1}{\xi -\rho ^{-2}}\right) \frac{d{y}}{d{\xi }} +  \frac{-\alpha (\alpha +1) \xi +h\rho ^{-2}}{4 \xi (\xi -1)(\xi -\rho ^{-2})} y(\xi ) = 0\label{eq:2}
\end{equation}
This is an equation of Fuchsian type with four regular singularities: $\xi=0, 1, \rho ^{-2}, \infty $. 
In Ref.\cite{Chou2012g}, we obtain power series expansion and integral form of Lame function of the first type in   Weierstrass's form by assume $\displaystyle{y(\xi ) = \sum_{n=0}^{\infty } c_n \xi^{n+\lambda }}$. Actually, there are three more types of Lame function in which are first one is $\displaystyle{y(\xi ) = (\xi-1)^{1/2}\sum_{n=0}^{\infty } b_n \xi^{n+\lambda }}$, second is $\displaystyle{y(\xi ) = (\xi- \rho ^{-2})^{1/2}\sum_{n=0}^{\infty } d_n \xi^{n+\lambda }}$ and third is  $\displaystyle{y(\xi ) = (\xi- 1)^{1/2}(\xi- \rho ^{-2})^{1/2}\sum_{n=0}^{\infty } e_n \xi^{n+\lambda }}$\cite{Wang1989}. We also have three term recurrence relation again. Then we obtain the analytic solution of all three cases by applying three term recurrence formula \cite{chou2012b}: if the time is permitted, I will publish these three cases of Lame equation in   Weierstrass's form.
\section{Generating function of Lame polynomial which makes $B_n$ term terminated in Weierstrass's form}
Let's investigate the generating function of the first and second kind of Lame polynomial as $B_n's$ term terminated at certain eigenvalue. 
\begin{lem}
The generating function of Jacobi polynomial using hypergeometric functions is defined by\footnote{In this paper Pochhammer symbol $(x)_n$ is used to represent the rising factorial: $(x)_n = \frac{\Gamma (x+n)}{\Gamma (x)}$.}
\begin{eqnarray}
&&\sum_{\alpha _0=0}^{\infty }\frac{(\gamma )_{\alpha _0}}{\alpha _0!} w^{\alpha _0} \;_2F_1(-\alpha _0, \alpha _0+A; \gamma; x) \label{eq:3}\\
&&= 2^{A -1}\frac{\left(1-w+\sqrt{w^2-2(1-2x)w+1}\right)^{1-\gamma } \left(1+w+\sqrt{w^2-2(1-2x)w+1}\right)^{\gamma -A}}{\sqrt{w^2-2(1-2x)w+1}} \;\;\mbox{where}\;|w|<1 \nonumber
\end{eqnarray}
\end{lem}
\begin{pf}
Jacobi polynomial $P_n^{(\alpha, \beta )}(x)$ can be written in terms of hypergeometric function using
\begin{equation}
_2F_1 (-n, n+\alpha +\beta +1; \alpha +1;x) = \frac{n!}{(\alpha +1)_n}P_n^{(\alpha, \beta )}(1-2x) \label{eq:20}
\end{equation}
And
\begin{equation}
P_n^{(\alpha, \beta )}(x) = \frac{\Gamma (n+\alpha +1)}{n! \Gamma (n+\alpha +\beta +1)} \sum_{m=0}^{n} \binom{n}{m} \frac{\Gamma (n+m+\alpha +\beta +1)}{\Gamma (m+\alpha +1)}\left( \frac{x-1}{2}\right)^m \label{eq:21}
\end{equation}
The generating function of the Jacobi polynomials is given by 
\begin{equation}
\sum_{n=0}^{\infty } P_n^{(\alpha, \beta )}(x) w^n = 2^{\alpha +\beta }\frac{\left(1-w+\sqrt{w^2-2xw+1}\right)^{-\alpha  } \left(1+w+\sqrt{w^2-2xw+1}\right)^{-\beta }}{\sqrt{w^2-2xw+1}} \label{eq:22}
\end{equation}
Replace n, $\alpha $ and $\beta $ by $\alpha _0$, $\gamma -1$ and $A-\gamma $ in (\ref{eq:20}), and acting the summation operator ${\displaystyle \sum_{\alpha _0=0}^{\infty }\frac{(\gamma )_{\alpha _0}}{\alpha _0!} w^{\alpha _0} }$ on the new (\ref{eq:20})
\begin{equation}
\sum_{\alpha _0=0}^{\infty }\frac{(\gamma )_{\alpha _0}}{\alpha _0!} w^{\alpha _0}\; _2F_1 (-\alpha _0, \alpha _0+ A; \gamma ;x) = \sum_{\alpha _0=0}^{\infty } P_n^{(\gamma -1, A-\gamma  )}(1-2x) w^{\alpha _0}\label{eq:23}
\end{equation}
Replace $\alpha $, $\beta $ and $x$ by $\gamma -1$, $A-\gamma $ and $1-2x$ in (\ref{eq:22}). As we take the new (\ref{eq:22}) into (\ref{eq:23}), we obtain (\ref{eq:3}).
\qed
\end{pf}
\begin{defn}
I define that
\begin{equation}
\begin{cases}
\displaystyle { s_{a,b}} = \begin{cases} \displaystyle {  s_a\cdot s_{a+1}\cdot s_{a+2}\cdots s_{b-2}\cdot s_{b-1}\cdot s_b}\;\;\mbox{where}\;a>b \cr
s_a \;\;\mbox{where}\;a=b\end{cases}
\cr
\cr
\displaystyle { \widetilde{w}_{i,j}}  = \begin{cases} \displaystyle { \frac{\widetilde{w}_{i+1,j}\; t_i u_i \left\{ 1+ (s_i+2\widetilde{w}_{i+1,j}(1-t_i)(1-u_i))s_i-(1+s_i)\sqrt{s_i^2-2(1-2\widetilde{w}_{i+1,j}(1-t_i)(1-u_i))s_i+1}\right\}}{2(1-\widetilde{w}_{i+1,j}(1-t_i)(1-u_i))^2 s_i}} \;\;\mbox{where}\;i<j \cr
\cr
\displaystyle { \frac{\eta t_i u_i \left\{ 1+ (s_{i,\infty }+2\eta(1-t_i)(1-u_i))s_{i,\infty }-(1+s_{i,\infty })\sqrt{s_{i,\infty }^2-2(1-2\eta (1-t_i)(1-u_i))s_{i,\infty }+1}\right\}}{2(1-\eta (1-t_i)(1-u_i))^2 s_{i,\infty }}} \;\;\mbox{where}\;i=j \end{cases}
\end{cases}\label{eq:4}
\end{equation}
\end{defn}
And we have
\begin{equation}
\sum_{\alpha _i = \alpha _j}^{\infty } r_i^{\alpha _i} = \frac{r_i^{\alpha _j}}{(1-r_i)}\label{eq:5}
\end{equation}
\begin{thm}
The general expression of the representation in the form of integral of Lame polynomial which makes $B_n$ term terminated in Weierstrass's form in Ref.\cite{Chou2012g} is given by\footnote{If we take $\alpha \geq -\frac{1}{2}$, $\alpha =  -2 (2\alpha _i+i +\lambda )-1 $ is not available any more. In this paper I consider $\alpha $ as arbitrary.}
\begin{eqnarray}
 y(\xi )&=& \sum_{n=0}^{\infty } y_{n}(\xi )= y_0(\xi )+y_1(\xi )+y_2(\xi )+y_3(\xi )+ \cdots \nonumber\\
&=& c_0 \xi ^{\lambda } \Bigg\{ \sum_{i_0=0}^{\alpha _0}\frac{(-\alpha _0)_{i_0}(\alpha _0+\frac{1}{4}+\lambda )_{i_0}}{(1+\frac{\lambda }{2})_{i_0}(\frac{3}{4}+ \frac{\lambda }{2})_{i_0}}  \eta ^{i_0}\nonumber\\
&&+ \sum_{n=1}^{\infty } \Bigg\{\prod _{k=0}^{n-1} \Bigg( \int_{0}^{1} dt_{n-k}\;t_{n-k}^{\frac{1}{2}(n-k-\frac{5}{2}+\lambda )} \int_{0}^{1} du_{n-k}\;u_{n-k}^{\frac{1}{2}(n-k-2+\lambda )} \nonumber\\
&&\times  \frac{1}{2\pi i}  \oint dv_{n-k} \frac{1}{v_{n-k}} \left( 1-\overleftrightarrow {w}_{n-k+1,n} v_{n-k}(1-t_{n-k})(1-u_{n-k})\right)^{-(n-k+\frac{1}{4}+\lambda )}\nonumber\\
&&\times \left(\frac{(v_{n-k}-1)}{v_{n-k}} \frac{1}{1-\overleftrightarrow {w}_{n-k+1,n}v_{n-k}(1-t_{n-k})(1-u_{n-k})}\right)^{\alpha _{n-k}}\nonumber\\
&&\times \Bigg( \overleftrightarrow {w}_{n-k,n}^{-\frac{1}{2}(n-k-1+\lambda )}\left(  \overleftrightarrow {w}_{n-k,n} \partial _{ \overleftrightarrow {w}_{n-k,n}}\right)^2 \overleftrightarrow {w}_{n-k,n}^{\frac{1}{2}(n-k-1+\lambda )} -\frac{h}{2^4(1+\rho ^2)}\Bigg) \Bigg)\nonumber\\
&&\times \sum_{i_0=0}^{\alpha _0}\frac{(-\alpha _0)_{i_0}(\alpha _0+\frac{1}{4}+\lambda )_{i_0}}{(1+\frac{\lambda }{2})_{i_0}(\frac{3}{4}+ \frac{\lambda }{2})_{i_0}}  \overleftrightarrow {w}_{1,n}^{i_0}\Bigg\} \mu ^n \Bigg\} \label{eq:6}
\end{eqnarray}
where
\begin{equation}
\begin{cases} 
\xi = sn^2(z,\rho ) \cr
 \mu = (1+\rho ^2) \xi \cr
 \eta = -\rho ^2 \xi^2  
\end{cases} \nonumber  
\end{equation}
\begin{equation}
\begin{cases} 
\alpha = 2( 2\alpha _i+ i+\lambda )\;\mbox{or}\; -2 (2\alpha _i+i +\lambda )-1\;\;\mbox{where}\; i,\alpha _i =0,1,2,\cdots \cr
\alpha _i\leq \alpha _j \;\;\mbox{only}\;\mbox{if}\;i\leq j\;\;\mbox{where}\;i,j \ =0,1,2,\cdots 
\end{cases}\nonumber
\end{equation}
And
\begin{equation}\overleftrightarrow {w}_{i,j}=
\begin{cases} \displaystyle {\frac{1}{(v_i-1)}\; \frac{\overleftrightarrow w_{i+1,j}v_i t_i u_i}{1- \overleftrightarrow w_{i+1,j} v_i (1-t_i)(1-u_i)}}\cr
\eta \;\;\mbox{only}\;\mbox{if}\; i>j
\end{cases}\nonumber 
\end{equation}
\end{thm}
Acting the summation operator $\displaystyle{ \sum_{\alpha _0 =0}^{\infty } \frac{(\gamma)_{\alpha _0}}{\alpha _0!} s_0^{\alpha _0} \prod _{n=1}^{\infty } \left\{ \sum_{ \alpha _n = \alpha _{n-1}}^{\infty } s_n^{\alpha _n }\right\}}$ on (\ref{eq:6}) where $|s_i|<1$ as $i=0,1,2,\cdots$ by using (\ref{eq:4}) and (\ref{eq:5}), 
\begin{thm}  
The general expression of the generating function of Lame polynomial which makes $B_n$ term terminated in Weierstrass's form about $\xi=0$ is given by
\begin{eqnarray}
&&\sum_{\alpha _0 =0}^{\infty } \frac{(\gamma)_{\alpha _0}}{\alpha _0!} s_0^{\alpha _0} \prod _{n=1}^{\infty } \left\{ \sum_{ \alpha _n = \alpha _{n-1}}^{\infty } s_n^{\alpha _n }\right\} y(\xi) \nonumber\\
&&= \prod_{l=1}^{\infty } \frac{1}{(1-s_{l,\infty })} \mathbf{\Upsilon}(\lambda; s_{0,\infty } ;\eta)\nonumber\\
&&+ \Bigg\{ \prod_{l=2}^{\infty } \frac{1}{(1-s_{l,\infty })} \int_{0}^{1} dt_1\;t_1^{\frac{1}{2}(-\frac{3}{2}+\lambda )} \int_{0}^{1} du_1\;u_1^{\frac{1}{2}(-1+\lambda )} \frac{\left( \frac{(1+s_{1,\infty })+\sqrt{s_{1,\infty }^2-2(1-2\eta (1-t_1)(1-u_1))s_{1,\infty }+1}}{2}\right)^{-(\frac{1}{4}+\lambda )}}{\sqrt{s_{1,\infty }^2-2(1-2\eta (1-t_1)(1-u_1))s_{1,\infty }+1}}\nonumber\\
&&\times  \Bigg( \widetilde{w}_{1,1}^{-\frac{\lambda }{2}}\left(  \widetilde{w}_{1,1} \partial _{ \widetilde{w}_{1,1}}\right)^2 \widetilde{w}_{1,1}^{\frac{\lambda }{2}} -\frac{h}{2^4 (1+\rho ^2)}\Bigg) \mathbf{\Upsilon}(\lambda ; s_0;\widetilde{w}_{1,1})\Bigg\} \mu \nonumber\\
&&+ \sum_{n=2}^{\infty } \Bigg\{ \prod_{l=n+1}^{\infty } \frac{1}{(1-s_{l,\infty })} \int_{0}^{1} dt_n\;t_n^{\frac{1}{2}(n-\frac{5}{2}+\lambda )} \int_{0}^{1} du_n\;u_n^{\frac{1}{2}(n-2+\lambda )} \frac{ \left( \frac{(1+s_{n,\infty })+\sqrt{s_{n,\infty }^2-2(1-2\eta (1-t_n)(1-u_n))s_{n,\infty }+1}}{2}\right)^{-(n-\frac{3}{4}+\lambda )} }{\sqrt{s_{n,\infty }^2-2(1-2\eta (1-t_n)(1-u_n))s_{n,\infty }+1}}\nonumber\\
&&\times \Bigg( \widetilde{w}_{n,n}^{-\frac{1}{2}(n-1+\lambda )}\left( \widetilde{w}_{n,n} \partial _{ \widetilde{w}_{n,n}}\right)^2 \widetilde{w}_{n,n}^{\frac{1}{2}(n-1+\lambda )}-\frac{h}{2^4 (1+\rho ^2)}\Bigg) \nonumber\\
&&\times \prod_{k=1}^{n-1} \Bigg\{ \int_{0}^{1} dt_{n-k}\;t_{n-k}^{\frac{1}{2}(n-k-\frac{5}{2}+\lambda )} \int_{0}^{1} du_{n-k} \;u_{n-k}^{\frac{1}{2}(n-k-2+\lambda )} \frac{ \left( \frac{(1+s_{n-k})+\sqrt{s_{n-k}^2-2(1-2\widetilde{w}_{n+1-k,n} (1-t_{n-k})(1-u_{n-k}))s_{n-k}+1}}{2}\right)^{-(n-k-\frac{3}{4}+\lambda )} }{\sqrt{s_{n-k}^2-2(1-2\widetilde{w}_{n+1-k,n} (1-t_{n-k})(1-u_{n-k}))s_{n-k}+1}}  \nonumber\\
&&\times \Bigg( \widetilde{w}_{n-k,n}^{-\frac{1}{2}(n-k-1+\lambda )}\left( \widetilde{w}_{n-k,n} \partial _{ \widetilde{w}_{n-k,n}}\right)^2 \widetilde{w}_{n-k,n}^{\frac{1}{2}(n-k-1+\lambda )}-\frac{h}{2^4 (1+\rho ^2)}\Bigg) \Bigg\} \mathbf{\Upsilon}(\lambda; s_0 ;\widetilde{w}_{1,n}) \Bigg\} \mu ^n\label{eq:8}
\end{eqnarray}
where
\begin{equation}
\begin{cases} 
{ \displaystyle \mathbf{\Upsilon}(\lambda; s_{0,\infty } ;\eta)= \sum_{\alpha _0 =0}^{\infty } \frac{(\gamma)_{\alpha _0}}{\alpha _0!} s_{0,\infty }^{\alpha _0} \Bigg\{ c_0 \xi^{\lambda } \sum_{i_0=0}^{\alpha _0} \frac{(-\alpha _0)_{i_0} (\alpha _0+\frac{1}{4}+\lambda )_{i_0}}{(1+\frac{\lambda }{2})_{i_0}(\frac{3}{4}+ \frac{\lambda }{2})_{i_0}} \eta ^{i_0} \Bigg\} }\cr
{ \displaystyle \mathbf{\Upsilon}(\lambda ; s_0;\widetilde{w}_{1,1}) = \sum_{\alpha _0 =0}^{\infty } \frac{(\gamma)_{\alpha _0}}{\alpha _0!} s_0^{\alpha _0}\Bigg\{c_0 \xi^{\lambda} \sum_{i_0=0}^{\alpha _0} \frac{(-\alpha _0)_{i_0} (\alpha _0+\frac{1}{4}+\lambda )_{i_0}}{(1+\frac{\lambda }{2})_{i_0}(\frac{3}{4}+\frac{\lambda }{2})_{i_0}} \widetilde{w}_{1,1} ^{i_0} \Bigg\} }\cr
{ \displaystyle \mathbf{\Upsilon}(\lambda; s_0 ;\widetilde{w}_{1,n}) = \sum_{\alpha _0 =0}^{\infty } \frac{(\gamma)_{\alpha _0}}{\alpha _0!} s_0^{\alpha _0}\Bigg\{c_0 \xi^{\lambda} \sum_{i_0=0}^{\alpha _0} \frac{(-\alpha _0)_{i_0} (\alpha _0+\frac{1}{4}+\lambda )_{i_0}}{(1+\frac{\lambda }{2})_{i_0}(\frac{3}{4}+\frac{\lambda }{2})_{i_0}} \widetilde{w}_{1,n} ^{i_0} \Bigg\}}
\end{cases}\nonumber 
\end{equation}

\end{thm}
\begin{pot} 
Acting the summation operator $\displaystyle{ \sum_{\alpha _0 =0}^{\infty } \frac{(\gamma)_{\alpha _0}}{\alpha _0!} s_0^{\alpha _0} \prod _{n=1}^{\infty } \left\{ \sum_{ \alpha _n = \alpha _{n-1}}^{\infty } s_n^{\alpha _n }\right\}}$ on the form of integral of Lame polynomial in Weierstrass's form which makes $B_n$ term terminated $y(\xi )$,
\begin{eqnarray}
&&\sum_{\alpha _0 =0}^{\infty } \frac{(\gamma)_{\alpha _0}}{\alpha _0!} s_0^{\alpha _0} \prod _{n=1}^{\infty } \left\{ \sum_{ \alpha _n = \alpha _{n-1}}^{\infty } s_n^{\alpha _n }\right\} y(\xi) \nonumber\\
&&= \sum_{\alpha _0 =0}^{\infty } \frac{(\gamma )_{\alpha _0}}{\alpha _0!} s_0^{\alpha _0} \prod _{n=1}^{\infty } \left\{ \sum_{ \alpha _n = \alpha _{n-1}}^{\infty } s_n^{\alpha _n }\right\} \left\{ y_0(\xi )+y_1(\xi )+y_2(\xi )+y_3(\xi )+ \cdots \right\} \label{eq:80}
\end{eqnarray}
According to (\ref{eq:6}), integral forms of sub-summation $y_0(\xi) $, $y_1(\xi)$, $y_2(\xi)$ and $y_3(\xi)$ are
\begin{subequations}
\begin{equation}
 y_0(\xi)= c_0 \xi^{\lambda } \sum_{i_0=0}^{\alpha _0} \frac{(-\alpha _0)_{i_0} (\alpha _0+ \frac{1}{4}+\lambda )_{i_0}}{(1+\frac{\lambda }{2})_{i_0}(\frac{3}{4}+ \frac{\lambda }{2})_{i_0}} \eta ^{i_0}\label{eq:24a}
\end{equation}
\begin{eqnarray}
 y_1(\xi) &=& \int_{0}^{1} dt_1\;t_1^{\frac{1}{2}(-\frac{3}{2}+\lambda) } \int_{0}^{1} du_1\;u_1^{\frac{1}{2}(-1+\lambda )}
 \frac{1}{2\pi i} \oint dv_1 \;\frac{1}{v_1} (1-\eta (1-t_1)(1-u_1)v_1)^{-(\frac{5}{4}+\lambda )} \nonumber\\
&&\times \left( \frac{v_1-1}{v_1} \frac{1}{1-\eta (1-t_1)(1-u_1)v_1}\right)^{\alpha _1} \Bigg( \overleftrightarrow {w}_{1,1}^{-\frac{\lambda }{2}}\left(  \overleftrightarrow {w}_{1,1} \partial _{ \overleftrightarrow {w}_{1,1}}\right) \overleftrightarrow {w}_{1,1}^{\frac{\lambda }{2}} -\frac{h}{2^4 (1+\rho ^2)}\Bigg) \nonumber\\
&&\times \left\{ c_0 \xi^{\lambda } \sum_{i_0=0}^{\alpha _0} \frac{(-\alpha _0)_{i_0} (\alpha _0+ \frac{1}{4}+\lambda )_{i_0}}{(1+\frac{\lambda }{2})_{i_0}(\frac{3}{4}+ \frac{\lambda }{2})_{i_0}} \overleftrightarrow {w}_{1,1} ^{i_0} \right\} \mu  \label{eq:24b}
\end{eqnarray}
\begin{eqnarray}
 y_2(\xi) &=& \int_{0}^{1} dt_2\;t_2^{\frac{1}{2}(-\frac{1}{2}+\lambda) } \int_{0}^{1} du_2\;u_2^{\frac{\lambda }{2}}
 \frac{1}{2\pi i} \oint dv_2 \;\frac{1}{v_2} (1-\eta (1-t_2)(1-u_2)v_2)^{-(\frac{9}{4}+\lambda )} \nonumber\\
&&\times \left( \frac{v_2-1}{v_2} \frac{1}{1-\eta (1-t_2)(1-u_2)v_2}\right)^{\alpha _2} \Bigg( \overleftrightarrow {w}_{2,2}^{-\frac{1}{2}(1+\lambda )}\left(  \overleftrightarrow {w}_{2,2} \partial _{ \overleftrightarrow {w}_{2,2}}\right) \overleftrightarrow {w}_{2,2}^{\frac{1}{2}(1+\lambda )} -\frac{h}{2^4 (1+\rho ^2)}\Bigg) \nonumber\\
&&\times \int_{0}^{1} dt_1\;t_1^{\frac{1}{2}(-\frac{3}{2}+\lambda) } \int_{0}^{1} du_1\;u_1^{\frac{1}{2}(-1+\lambda )}
 \frac{1}{2\pi i} \oint dv_1 \;\frac{1}{v_1} (1-\overleftrightarrow {w}_{2,2} (1-t_1)(1-u_1)v_1)^{-(\frac{5}{4}+\lambda )} \nonumber\\
&&\times \left( \frac{v_1-1}{v_1} \frac{1}{1-\overleftrightarrow {w}_{2,2}(1-t_1)(1-u_1)v_1}\right)^{\alpha _1} \Bigg( \overleftrightarrow {w}_{1,2}^{-\frac{\lambda }{2}}\left(  \overleftrightarrow {w}_{1,2} \partial _{ \overleftrightarrow {w}_{1,2}}\right) \overleftrightarrow {w}_{1,2}^{\frac{\lambda }{2}} -\frac{h}{2^4 (1+\rho ^2)}\Bigg) \nonumber\\
&&\times \left\{ c_0 \xi^{\lambda } \sum_{i_0=0}^{\alpha _0} \frac{(-\alpha _0)_{i_0} (\alpha _0+ \frac{1}{4}+\lambda )_{i_0}}{(1+\frac{\lambda }{2})_{i_0}(\frac{3}{4}+ \frac{\lambda }{2})_{i_0}} \overleftrightarrow {w}_{1,2} ^{i_0} \right\} \mu ^2 \label{eq:24c}
\end{eqnarray}
\begin{eqnarray}
 y_3(\xi) &=& \int_{0}^{1} dt_3\;t_3^{\frac{1}{2}(\frac{1}{2}+\lambda) } \int_{0}^{1} du_3\;u_3^{\frac{1}{2}(1+\lambda )}
 \frac{1}{2\pi i} \oint dv_3 \;\frac{1}{v_3} (1-\eta (1-t_3)(1-u_3)v_3)^{-(\frac{13}{4}+\lambda )} \nonumber\\
&&\times \left( \frac{v_3-1}{v_3} \frac{1}{1-\eta (1-t_3)(1-u_3)v_3}\right)^{\alpha _3} \Bigg( \overleftrightarrow {w}_{3,3}^{-\frac{1}{2}(2+\lambda )}\left(  \overleftrightarrow {w}_{3,3} \partial _{ \overleftrightarrow {w}_{3,3}}\right) \overleftrightarrow {w}_{3,3}^{\frac{1}{2}(2+\lambda )} -\frac{h}{2^4 (1+\rho ^2)}\Bigg) \nonumber\\
&&\times \int_{0}^{1} dt_2\;t_2^{\frac{1}{2}(-\frac{1}{2}+\lambda) } \int_{0}^{1} du_2\;u_2^{\frac{\lambda }{2}}
 \frac{1}{2\pi i} \oint dv_2 \;\frac{1}{v_2} (1-\overleftrightarrow {w}_{3,3} (1-t_2)(1-u_2)v_2)^{-(\frac{9}{4}+\lambda )} \nonumber\\
&&\times \left( \frac{v_2-1}{v_2} \frac{1}{1-\overleftrightarrow {w}_{3,3} (1-t_2)(1-u_2)v_2}\right)^{\alpha _2} \Bigg( \overleftrightarrow {w}_{2,3}^{-\frac{1}{2}(1+\lambda )}\left(  \overleftrightarrow {w}_{2,3} \partial _{ \overleftrightarrow {w}_{2,3}}\right) \overleftrightarrow {w}_{2,3}^{\frac{1}{2}(1+\lambda )} -\frac{h}{2^4 (1+\rho ^2)}\Bigg) \nonumber\\
&&\times \int_{0}^{1} dt_1\;t_1^{\frac{1}{2}(-\frac{3}{2}+\lambda) } \int_{0}^{1} du_1\;u_1^{\frac{1}{2}(-1+\lambda )}
 \frac{1}{2\pi i} \oint dv_1 \;\frac{1}{v_1} (1-\overleftrightarrow {w}_{2,3} (1-t_1)(1-u_1)v_1)^{-(\frac{5}{4}+\lambda )} \nonumber\\
&&\times \left( \frac{v_1-1}{v_1} \frac{1}{1-\overleftrightarrow {w}_{2,3}(1-t_1)(1-u_1)v_1}\right)^{\alpha _1} \Bigg( \overleftrightarrow {w}_{1,3}^{-\frac{\lambda }{2}}\left(  \overleftrightarrow {w}_{1,3} \partial _{ \overleftrightarrow {w}_{1,3}}\right) \overleftrightarrow {w}_{1,3}^{\frac{\lambda }{2}} -\frac{h}{2^4 (1+\rho ^2)}\Bigg) \nonumber\\
&&\times \left\{ c_0 \xi^{\lambda } \sum_{i_0=0}^{\alpha _0} \frac{(-\alpha _0)_{i_0} (\alpha _0+ \frac{1}{4}+\lambda )_{i_0}}{(1+\frac{\lambda }{2})_{i_0}(\frac{3}{4}+ \frac{\lambda }{2})_{i_0}} \overleftrightarrow {w}_{1,3} ^{i_0} \right\} \mu ^3 \label{eq:24d}
\end{eqnarray}
\end{subequations}
Acting the summation operator $\displaystyle{ \sum_{\alpha _0 =0}^{\infty } \frac{(\gamma)_{\alpha _0}}{\alpha _0!} s_0^{\alpha _0} \prod _{n=1}^{\infty } \left\{ \sum_{ \alpha _n = \alpha _{n-1}}^{\infty } s_n^{\alpha _n }\right\}}$ on (\ref{eq:24a}),
\begin{eqnarray}
&&\sum_{\alpha _0 =0}^{\infty } \frac{(\gamma )_{\alpha _0}}{\alpha _0!} s_0^{\alpha _0} \prod _{n=1}^{\infty } \left\{ \sum_{ \alpha _n = \alpha _{n-1}}^{\infty } s_n^{\alpha _n }\right\} y_0(\xi) \nonumber\\
&&= \prod_{l=1}^{\infty } \frac{1}{(1-s_{l,\infty })} \sum_{\alpha _0 =0}^{\infty } \frac{(\gamma)_{\alpha _0}}{\alpha _0!} s_{0,\infty }^{\alpha _0} \left\{ c_0 \xi^{\lambda } \sum_{i_0=0}^{\alpha _0} \frac{(-\alpha _0)_{i_0} (\alpha _0+ \frac{1}{4}+\lambda )_{i_0}}{(1+\frac{\lambda }{2})_{i_0}(\frac{3}{4}+ \frac{\lambda }{2})_{i_0}} \eta ^{i_0} \right\}\label{eq:25}
\end{eqnarray}
Acting the summation operator $\displaystyle{ \sum_{\alpha _0 =0}^{\infty } \frac{(\gamma)_{\alpha _0}}{\alpha _0!} s_0^{\alpha _0} \prod _{n=1}^{\infty } \left\{ \sum_{ \alpha _n = \alpha _{n-1}}^{\infty } s_n^{\alpha _n }\right\}}$ on (\ref{eq:24b}),
\begin{eqnarray}
&&\sum_{\alpha _0 =0}^{\infty } \frac{(\gamma)_{\alpha _0}}{\alpha _0!} s_0^{\alpha _0} \prod _{n=1}^{\infty } \left\{ \sum_{ \alpha _n = \alpha _{n-1}}^{\infty } s_n^{\alpha _n }\right\} y_1(\xi) \nonumber\\
&&= \prod_{l=2}^{\infty } \frac{1}{(1-s_{l,\infty })} \int_{0}^{1} dt_1\;t_1^{\frac{1}{2}(-\frac{3}{2}+\lambda) } \int_{0}^{1} du_1\;u_1^{\frac{1}{2}(-1+\lambda )}
 \frac{1}{2\pi i} \oint dv_1 \;\frac{1}{v_1} (1-\eta (1-t_1)(1-u_1)v_1)^{-(\frac{5}{4}+\lambda )} \nonumber\\
&&\times \sum_{\alpha _1 =\alpha _0}^{\infty }\left( \frac{v_1-1}{v_1} \frac{s_{1,\infty }}{1-\eta (1-t_1)(1-u_1)v_1}\right)^{\alpha _1} \Bigg( \overleftrightarrow {w}_{1,1}^{-\frac{\lambda }{2}}\left(  \overleftrightarrow {w}_{1,1} \partial _{ \overleftrightarrow {w}_{1,1}}\right) \overleftrightarrow {w}_{1,1}^{\frac{\lambda }{2}} -\frac{h}{2^4 (1+\rho ^2)}\Bigg) \nonumber\\
&&\times  \sum_{\alpha _0 =0}^{\infty } \frac{(\gamma )_{\alpha _0}}{\alpha _0!}s_0^{\alpha _0}\left\{ c_0 \xi^{\lambda } \sum_{i_0=0}^{\alpha _0} \frac{(-\alpha _0)_{i_0} (\alpha _0+ \frac{1}{4}+\lambda )_{i_0}}{(1+\frac{\lambda }{2})_{i_0}(\frac{3}{4}+ \frac{\lambda }{2})_{i_0}} \overleftrightarrow {w}_{1,1} ^{i_0} \right\} \mu \label{eq:26}
\end{eqnarray}
Replace $\alpha _i$, $\alpha _j$ and $r_i$ by $\alpha _1$, $\alpha _0$ and ${ \displaystyle \frac{v_1-1}{v_1} \frac{s_{1,\infty }}{1-\eta (1-t_1)(1-u_1)v_1}}$ in (\ref{eq:5}). Take the new (\ref{eq:5}) into (\ref{eq:26}).
\begin{eqnarray}
&&\sum_{\alpha _0 =0}^{\infty } \frac{(\gamma)_{\alpha _0}}{\alpha _0!} s_0^{\alpha _0} \prod _{n=1}^{\infty } \left\{ \sum_{ \alpha _n = \alpha _{n-1}}^{\infty } s_n^{\alpha _n }\right\} y_1(\xi) \nonumber\\
&&= \prod_{l=2}^{\infty } \frac{1}{(1-s_{l,\infty })} \int_{0}^{1} dt_1\;t_1^{\frac{1}{2}(-\frac{3}{2}+\lambda) } \int_{0}^{1} du_1\;u_1^{\frac{1}{2}(-1+\lambda )}
 \frac{1}{2\pi i} \oint dv_1 \;\frac{-(1-\eta (1-t_1)(1-u_1)v_1)^{-(\frac{1}{4}+\lambda )} }{\eta (1-t_1)(1-u_1)v_1^2+ (s_{1,\infty }-1)v_1-s_{1,\infty } } \nonumber\\
&&\times \Bigg( \overleftrightarrow {w}_{1,1}^{-\frac{\lambda }{2}}\left(  \overleftrightarrow {w}_{1,1} \partial _{ \overleftrightarrow {w}_{1,1}}\right) \overleftrightarrow {w}_{1,1}^{\frac{\lambda }{2}} -\frac{h}{2^4 (1+\rho ^2)}\Bigg) \nonumber\\
&&\times  \sum_{\alpha _0 =0}^{\infty } \frac{(\gamma )_{\alpha _0}}{\alpha _0!} \left( \frac{v_1-1}{v_1} \frac{s_{0,\infty }}{1-\eta (1-t_1)(1-u_1)v_1}\right)^{\alpha _0}\left\{ c_0 \xi^{\lambda } \sum_{i_0=0}^{\alpha _0} \frac{(-\alpha _0)_{i_0} (\alpha _0+ \frac{1}{4}+\lambda )_{i_0}}{(1+\frac{\lambda }{2})_{i_0}(\frac{3}{4}+ \frac{\lambda }{2})_{i_0}} \overleftrightarrow {w}_{1,1} ^{i_0} \right\} \mu \label{eq:27}
\end{eqnarray}
By using Cauchy's integral formula, the contour integrand has poles at
\begin{equation}
 v_1= \frac{1-s_{1,\infty }-\sqrt{(1-s_{1,\infty })^2+4\eta (1-t_1)(1-u_1)s_{1,\infty }}}{2\eta (1-t_1)(1-u_1)}  \;\;\mbox{or}\;\frac{1-s_{1,\infty }+\sqrt{(1-s_{1,\infty })^2+4\eta (1-t_1)(1-u_1)s_{1,\infty }}}{2\eta (1-t_1)(1-u_1)} \nonumber
\end{equation}
and ${ \displaystyle \frac{1-s_{1,\infty }-\sqrt{(1-s_{1,\infty })^2+4\eta (1-t_1)(1-u_1)s_{1,\infty }}}{2\eta (1-t_1)(1-u_1)}}$ is only inside the unit circle. As we compute the residue there in (\ref{eq:27}) we obtain
\begin{eqnarray}
&&\sum_{\alpha _0 =0}^{\infty } \frac{(\gamma)_{\alpha _0}}{\alpha _0!} s_0^{\alpha _0} \prod _{n=1}^{\infty } \left\{ \sum_{ \alpha _n = \alpha _{n-1}}^{\infty } s_n^{\alpha _n }\right\} y_1(\xi) \label{eq:28}\\
&&= \prod_{l=2}^{\infty } \frac{1}{(1-s_{l,\infty })} \int_{0}^{1} dt_1\;t_1^{\frac{1}{2}(-\frac{3}{2}+\lambda) } \int_{0}^{1} du_1\;u_1^{\frac{1}{2}(-1+\lambda )}
 \frac{\left(\frac{1+s_{1,\infty }+\sqrt{s_{1,\infty }^2-2(1-2\eta (1-t_1)(1-u_1))s_{1,\infty }+1}}{2}\right)^{-(\frac{1}{4}+\lambda )}}{\sqrt{s_{1,\infty }^2-2(1-2\eta (1-t_1)(1-u_1))s_{1,\infty }+1}} \nonumber\\
&&\times \Bigg( \widetilde{w}_{1,1}^{-\frac{\lambda }{2}}\left(  \widetilde{w}_{1,1} \partial _{ \widetilde{w}_{1,1}}\right) \widetilde{w}_{1,1}^{\frac{\lambda }{2}} -\frac{h}{2^4 (1+\rho ^2)}\Bigg) 
  \sum_{\alpha _0 =0}^{\infty } \frac{(\gamma )_{\alpha _0}}{\alpha _0!} s_0^{\alpha _0}\left\{ c_0 \xi^{\lambda } \sum_{i_0=0}^{\alpha _0} \frac{(-\alpha _0)_{i_0} (\alpha _0+ \frac{1}{4}+\lambda )_{i_0}}{(1+\frac{\lambda }{2})_{i_0}(\frac{3}{4}+ \frac{\lambda }{2})_{i_0}} \widetilde{w}_{1,1} ^{i_0} \right\} \mu \nonumber
\end{eqnarray}
where
\begin{eqnarray}
\widetilde{w}_{1,1} &=& \frac{v_1}{(v_1-1)}\; \frac{\eta t_1 u_1}{1- \eta v_1 (1-t_1)(1-u_1)}\Bigg|_{\Large v_1=\frac{1-s_{1,\infty }-\sqrt{(1-s_{1,\infty })^2+4\eta (1-t_1)(1-u_1)s_{1,\infty }}}{2\eta (1-t_1)(1-u_1)}\normalsize}\nonumber\\
&=& \frac{\eta t_1 u_1 \left\{ 1+ (s_{1,\infty }+2\eta(1-t_1)(1-u_1) )s_{1,\infty }-(1+s_{1,\infty })\sqrt{s_{1,\infty }^2-2(1-2\eta (1-t_1)(1-u_1))s_{1,\infty }+1}\right\}}{2(1-\eta (1-t_1)(1-u_1))^2 s_{1,\infty }}\nonumber
\end{eqnarray}
Acting the summation operator $\displaystyle{ \sum_{\alpha _0 =0}^{\infty } \frac{(\gamma)_{\alpha _0}}{\alpha _0!} s_0^{\alpha _0} \prod _{n=1}^{\infty } \left\{ \sum_{ \alpha _n = \alpha _{n-1}}^{\infty } s_n^{\alpha _n }\right\}}$ on (\ref{eq:24c}),
\begin{eqnarray}
&&\sum_{\alpha _0 =0}^{\infty } \frac{(\gamma)_{\alpha _0}}{\alpha _0!} s_0^{\alpha _0} \prod _{n=1}^{\infty } \left\{ \sum_{ \alpha _n = \alpha _{n-1}}^{\infty } s_n^{\alpha _n }\right\} y_2(\xi) \nonumber\\
&&= \prod_{l=3}^{\infty } \frac{1}{(1-s_{l,\infty })} \int_{0}^{1} dt_2\;t_2^{\frac{1}{2}(-\frac{1}{2}+\lambda) } \int_{0}^{1} du_2\;u_2^{\frac{\lambda }{2}}
 \frac{1}{2\pi i} \oint dv_2 \;\frac{1}{v_2} (1-\eta (1-t_2)(1-u_2)v_2)^{-(\frac{9}{4}+\lambda )} \nonumber\\
&&\times \sum_{\alpha _2 =\alpha _1}^{\infty }\left( \frac{v_2-1}{v_2} \frac{s_{2,\infty }}{1-\eta (1-t_2)(1-u_2)v_2}\right)^{\alpha _2} \Bigg( \overleftrightarrow {w}_{2,2}^{-\frac{1}{2}(1+\lambda )}\left(  \overleftrightarrow {w}_{2,2} \partial _{ \overleftrightarrow {w}_{2,2}}\right) \overleftrightarrow {w}_{2,2}^{\frac{1}{2}(1+\lambda )} -\frac{h}{2^4 (1+\rho ^2)}\Bigg) \nonumber\\
&&\times \int_{0}^{1} dt_1\;t_1^{\frac{1}{2}(-\frac{3}{2}+\lambda) } \int_{0}^{1} du_1\;u_1^{\frac{1}{2}(-1+\lambda )}
 \frac{1}{2\pi i} \oint dv_1 \;\frac{1}{v_1} (1-\overleftrightarrow {w}_{2,2} (1-t_1)(1-u_1)v_1)^{-(\frac{5}{4}+\lambda )} \nonumber\\
&&\times \sum_{\alpha _1 =\alpha _0}^{\infty }\left( \frac{v_1-1}{v_1} \frac{s_1}{1-\overleftrightarrow {w}_{2,2}(1-t_1)(1-u_1)v_1}\right)^{\alpha _1} \Bigg(  \overleftrightarrow {w}_{1,2}^{-\frac{\lambda }{2}}\left(  \overleftrightarrow {w}_{1,2} \partial _{ \overleftrightarrow {w}_{1,2}}\right) \overleftrightarrow {w}_{1,2}^{\frac{\lambda }{2}} -\frac{h}{2^4 (1+\rho ^2)}\Bigg) \nonumber\\
&&\times  \sum_{\alpha _0 =0}^{\infty } \frac{(\gamma )_{\alpha _0}}{\alpha _0!}s_0^{\alpha _0}\left\{ c_0 \xi^{\lambda } \sum_{i_0=0}^{\alpha _0} \frac{(-\alpha _0)_{i_0} (\alpha _0+ \frac{1}{4}+\lambda )_{i_0}}{(1+\frac{\lambda }{2})_{i_0}(\frac{3}{4}+ \frac{\lambda }{2})_{i_0}} \overleftrightarrow {w}_{1,2} ^{i_0} \right\} \mu^2 \label{eq:29}
\end{eqnarray}
Replace $\alpha _i$, $\alpha _j$ and $r_i$ by $\alpha _2$, $\alpha _1$ and ${ \displaystyle \frac{v_2-1}{v_2} \frac{s_{2,\infty }}{1-\eta (1-t_2)(1-u_2)v_2}}$ in (\ref{eq:5}). Take the new (\ref{eq:5}) into (\ref{eq:29}).
\begin{eqnarray}
&&\sum_{\alpha _0 =0}^{\infty } \frac{(\gamma)_{\alpha _0}}{\alpha _0!} s_0^{\alpha _0} \prod _{n=1}^{\infty } \left\{ \sum_{ \alpha _n = \alpha _{n-1}}^{\infty } s_n^{\alpha _n }\right\} y_2(\xi) \nonumber\\
&&= \prod_{l=3}^{\infty } \frac{1}{(1-s_{l,\infty })} \int_{0}^{1} dt_2\;t_2^{\frac{1}{2}(-\frac{1}{2}+\lambda) } \int_{0}^{1} du_2\;u_2^{\frac{\lambda }{2}}
 \frac{1}{2\pi i} \oint dv_2 \;\frac{-\left(1-\eta (1-t_2)(1-u_2)v_2\right)^{-(\frac{5}{4}+\lambda )}}{\eta (1-t_2)(1-u_2)v_2^2+ (s_{2,\infty }-1)v_2-s_{2,\infty } } \nonumber\\
&&\times  \Bigg( \overleftrightarrow {w}_{2,2}^{-\frac{1}{2}(1+\lambda )}\left(  \overleftrightarrow {w}_{2,2} \partial _{ \overleftrightarrow {w}_{2,2}}\right) \overleftrightarrow {w}_{2,2}^{\frac{1}{2}(1+\lambda )} -\frac{h}{2^4 (1+\rho ^2)}\Bigg) \nonumber\\
&&\times \int_{0}^{1} dt_1\;t_1^{\frac{1}{2}(-\frac{3}{2}+\lambda) } \int_{0}^{1} du_1\;u_1^{\frac{1}{2}(-1+\lambda )}
 \frac{1}{2\pi i} \oint dv_1 \;\frac{1}{v_1} \left(1-\overleftrightarrow {w}_{2,2} (1-t_1)(1-u_1)v_1\right)^{-(\frac{5}{4}+\lambda )} \nonumber\\
&&\times \sum_{\alpha _1 =\alpha _0}^{\infty }\left( \frac{v_2-1}{v_2} \frac{s_{1,\infty }}{1-\eta (1-t_2)(1-u_2)v_2} \frac{v_1-1}{v_1}\frac{1}{1-\overleftrightarrow {w}_{2,2}(1-t_1)(1-u_1)v_1}\right)^{\alpha _1} \nonumber\\
&&\times \Bigg( \overleftrightarrow {w}_{1,2}^{-\frac{\lambda }{2}}\left(  \overleftrightarrow {w}_{1,2} \partial _{ \overleftrightarrow {w}_{1,2}}\right) \overleftrightarrow {w}_{1,2}^{\frac{\lambda }{2}} -\frac{h}{2^4 (1+\rho ^2)}\Bigg) \nonumber\\
&&\times  \sum_{\alpha _0 =0}^{\infty } \frac{(\gamma )_{\alpha _0}}{\alpha _0!} s_0^{\alpha _0}\left\{ c_0 \xi^{\lambda } \sum_{i_0=0}^{\alpha _0} \frac{(-\alpha _0)_{i_0} (\alpha _0+ \frac{1}{4}+\lambda )_{i_0}}{(1+\frac{\lambda }{2})_{i_0}(\frac{3}{4}+ \frac{\lambda }{2})_{i_0}} \overleftrightarrow {w}_{1,2} ^{i_0} \right\} \mu^2 \label{eq:30}
\end{eqnarray}
By using Cauchy's integral formula, the contour integrand has poles at
\begin{equation}
 v_2= \frac{1-s_{2,\infty }-\sqrt{(1-s_{2,\infty })^2+4\eta (1-t_2)(1-u_2)s_{2,\infty }}}{2\eta (1-t_2)(1-u_2)}  \;\;\mbox{or}\;\frac{1-s_{2,\infty }+\sqrt{(1-s_{2,\infty })^2+4\eta (1-t_2)(1-u_2)s_{2,\infty }}}{2\eta (1-t_2)(1-u_2)} \nonumber
\end{equation}
and ${ \displaystyle\frac{1-s_{2,\infty }-\sqrt{(1-s_{2,\infty })^2+4\eta (1-t_2)(1-u_2)s_{2,\infty }}}{2\eta (1-t_2)(1-u_2)}}$ is only inside the unit circle. As we compute the residue there in (\ref{eq:30}) we obtain
\begin{eqnarray}
&&\sum_{\alpha _0 =0}^{\infty } \frac{(\gamma)_{\alpha _0}}{\alpha _0!} s_0^{\alpha _0} \prod _{n=1}^{\infty } \left\{ \sum_{ \alpha _n = \alpha _{n-1}}^{\infty } s_n^{\alpha _n }\right\} y_2(\xi) \nonumber\\
&&= \prod_{l=3}^{\infty } \frac{1}{(1-s_{l,\infty })} \int_{0}^{1} dt_2\;t_2^{\frac{1}{2}(-\frac{1}{2}+\lambda) } \int_{0}^{1} du_2\;u_2^{\frac{\lambda }{2}}
 \frac{\left(\frac{1+s_{2,\infty }+\sqrt{s_{2,\infty }^2-2(1-2\eta (1-t_2)(1-u_2))s_{2,\infty }+1}}{2}\right)^{-(\frac{5}{4}+\lambda )}}{\sqrt{s_{2,\infty }^2-2(1-2\eta (1-t_2)(1-u_2))s_{2,\infty }+1}} \nonumber\\
&&\times  \Bigg( \widetilde{w}_{2,2}^{-\frac{1}{2}(1+\lambda )}\left( \widetilde{w}_{2,2} \partial _{ \widetilde{w}_{2,2}}\right) \widetilde{w}_{2,2}^{\frac{1}{2}(1+\lambda )} -\frac{h}{2^4 (1+\rho ^2)}\Bigg) \nonumber\\
&&\times \int_{0}^{1} dt_1\;t_1^{\frac{1}{2}(-\frac{3}{2}+\lambda) } \int_{0}^{1} du_1\;u_1^{\frac{1}{2}(-1+\lambda )}
 \frac{1}{2\pi i} \oint dv_1 \;\frac{1}{v_1} \left(1-\widetilde{w}_{2,2} (1-t_1)(1-u_1)v_1\right)^{-(\frac{5}{4}+\lambda )} \nonumber\\
&&\times \sum_{\alpha _1 =\alpha _0}^{\infty }\left( \frac{v_1-1}{v_1}\frac{s_1}{1-\widetilde{w}_{2,2}(1-t_1)(1-u_1)v_1}\right)^{\alpha _1}
 \Bigg( \ddot{w}_{1,2}^{-\frac{\lambda }{2}}\left(  \ddot{w}_{1,2} \partial _{ \ddot{w}_{1,2}}\right) \ddot{w}_{1,2}^{\frac{\lambda }{2}} -\frac{h}{2^4 (1+\rho ^2)}\Bigg) \nonumber\\
&&\times  \sum_{\alpha _0 =0}^{\infty } \frac{(\gamma )_{\alpha _0}}{\alpha _0!} s_0^{\alpha _0}\left\{ c_0 \xi^{\lambda } \sum_{i_0=0}^{\alpha _0} \frac{(-\alpha _0)_{i_0} (\alpha _0+ \frac{1}{4}+\lambda )_{i_0}}{(1+\frac{\lambda }{2})_{i_0}(\frac{3}{4}+ \frac{\lambda }{2})_{i_0}} \ddot{w}_{1,2} ^{i_0} \right\} \mu^2 \label{eq:31}
\end{eqnarray}
where
\begin{eqnarray}
\widetilde{w}_{2,2} &=& \frac{v_2}{(v_2-1)}\; \frac{\eta t_2 u_2}{1- \eta v_2 (1-t_2)(1-u_2)}\Bigg|_{\Large v_2=\frac{1-s_{2,\infty }-\sqrt{(1-s_{2,\infty })^2+4\eta (1-t_2)(1-u_2)s_{2,\infty }}}{2\eta (1-t_2)(1-u_2)}\normalsize}\nonumber\\
&=& \frac{\eta t_2 u_2 \left\{ 1+ (s_{2,\infty }+2\eta(1-t_2)(1-u_2) )s_{2,\infty }-(1+s_{2,\infty })\sqrt{s_{2,\infty }^2-2(1-2\eta (1-t_2)(1-u_2))s_{2,\infty }+1}\right\}}{2(1-\eta (1-t_2)(1-u_2))^2 s_{2,\infty }}\nonumber
\end{eqnarray}
and
\begin{equation}
\ddot{w}_{1,2} = \frac{v_1}{(v_1-1)}\; \frac{\widetilde{w}_{2,2} t_1 u_1}{1- \widetilde{w}_{2,2}v_1 (1-t_1)(1-u_1)}\nonumber
\end{equation}
Replace $\alpha _i$, $\alpha _j$ and $r_i$ by $\alpha _1$, $\alpha _0$ and ${ \displaystyle \frac{v_1-1}{v_1}\frac{s_1}{1-\widetilde{w}_{2,2}(1-t_1)(1-u_1)v_1}}$ in (\ref{eq:5}). Take the new (\ref{eq:5}) into (\ref{eq:31}).
\begin{eqnarray}
&&\sum_{\alpha _0 =0}^{\infty } \frac{(\gamma)_{\alpha _0}}{\alpha _0!} s_0^{\alpha _0} \prod _{n=1}^{\infty } \left\{ \sum_{ \alpha _n = \alpha _{n-1}}^{\infty } s_n^{\alpha _n }\right\} y_2(\xi ) \nonumber\\
&&= \prod_{l=3}^{\infty } \frac{1}{(1-s_{l,\infty })} \int_{0}^{1} dt_2\;t_2^{\frac{1}{2}(-\frac{1}{2}+\lambda) } \int_{0}^{1} du_2\;u_2^{\frac{\lambda }{2}}
 \frac{\left(\frac{1+s_{2,\infty }+\sqrt{s_{2,\infty }^2-2(1-2\eta (1-t_2)(1-u_2))s_{2,\infty }+1}}{2}\right)^{-(\frac{5}{4}+\lambda )}}{\sqrt{s_{2,\infty }^2-2(1-2\eta (1-t_2)(1-u_2))s_{2,\infty }+1}} \nonumber\\
&&\times  \Bigg( \widetilde{w}_{2,2}^{-\frac{1}{2}(1+\lambda )}\left( \widetilde{w}_{2,2} \partial _{ \widetilde{w}_{2,2}}\right) \widetilde{w}_{2,2}^{\frac{1}{2}(1+\lambda )} -\frac{h}{2^4 (1+\rho ^2)}\Bigg) \nonumber\\
&&\times \int_{0}^{1} dt_1\;t_1^{\frac{1}{2}(-\frac{3}{2}+\lambda) } \int_{0}^{1} du_1\;u_1^{\frac{1}{2}(-1+\lambda )}
 \frac{1}{2\pi i} \oint dv_1 \;\frac{-\left(1-\widetilde{w}_{2,2} (1-t_1)(1-u_1)v_1\right)^{-(\frac{1}{4}+\lambda )}}{\widetilde{w}_{2,2} (1-t_1)(1-u_1)v_1^2+(s_1-1)v_1-s_1} \nonumber\\
&&\times \Bigg( \ddot{w}_{1,2}^{-\frac{\lambda }{2}}\left(  \ddot{w}_{1,2} \partial _{ \ddot{w}_{1,2}}\right) \ddot{w}_{1,2}^{\frac{\lambda }{2}} -\frac{h}{2^4 (1+\rho ^2)}\Bigg) \label{eq:32}\\
&&\times  \sum_{\alpha _0 =0}^{\infty } \frac{(\gamma )_{\alpha _0}}{\alpha _0!} \left( \frac{v_1-1}{v_1}\frac{s_{0,1}}{1-\widetilde{w}_{2,2}(1-t_1)(1-u_1)v_1}\right)^{\alpha _0} \left\{ c_0 \xi^{\lambda } \sum_{i_0=0}^{\alpha _0} \frac{(-\alpha _0)_{i_0} (\alpha _0+ \frac{1}{4}+\lambda )_{i_0}}{(1+\frac{\lambda }{2})_{i_0}(\frac{3}{4}+ \frac{\lambda }{2})_{i_0}} \ddot{w}_{1,2} ^{i_0} \right\} \mu^2 \nonumber
\end{eqnarray}
By using Cauchy's integral formula, the contour integrand has poles at
\begin{equation}
 v_1= \frac{1-s_1-\sqrt{(1-s_1)^2+4\widetilde{w}_{2,2} (1-t_1)(1-u_1)s_1}}{2\widetilde{w}_{2,2} (1-t_1)(1-u_1)}  \;\;\mbox{or}\;\frac{1-s_1+\sqrt{(1-s_1)^2+4\widetilde{w}_{2,2} (1-t_1)(1-u_1)s_1}}{2\widetilde{w}_{2,2} (1-t_1)(1-u_1)}  \nonumber
\end{equation}
and ${ \displaystyle \frac{1-s_1-\sqrt{(1-s_1)^2+4\widetilde{w}_{2,2} (1-t_1)(1-u_1)s_1}}{2\widetilde{w}_{2,2} (1-t_1)(1-u_1)}}$ is only inside the unit circle. As we compute the residue there in (\ref{eq:32}) we obtain
\begin{eqnarray}
&&\sum_{\alpha _0 =0}^{\infty } \frac{(\gamma)_{\alpha _0}}{\alpha _0!} s_0^{\alpha _0} \prod _{n=1}^{\infty } \left\{ \sum_{ \alpha _n = \alpha _{n-1}}^{\infty } s_n^{\alpha _n }\right\} y_2(\xi ) \nonumber\\
&&= \prod_{l=3}^{\infty } \frac{1}{(1-s_{l,\infty })} \int_{0}^{1} dt_2\;t_2^{\frac{1}{2}(-\frac{1}{2}+\lambda) } \int_{0}^{1} du_2\;u_2^{\frac{\lambda }{2}}
 \frac{\left(\frac{1+s_{2,\infty }+\sqrt{s_{2,\infty }^2-2(1-2\eta (1-t_2)(1-u_2))s_{2,\infty }+1}}{2}\right)^{-(\frac{5}{4}+\lambda )}}{\sqrt{s_{2,\infty }^2-2(1-2\eta (1-t_2)(1-u_2))s_{2,\infty }+1}} \nonumber\\
&&\times  \Bigg( \widetilde{w}_{2,2}^{-\frac{1}{2}(1+\lambda )}\left( \widetilde{w}_{2,2} \partial _{ \widetilde{w}_{2,2}}\right) \widetilde{w}_{2,2}^{\frac{1}{2}(1+\lambda )} -\frac{h}{2^4 (1+\rho ^2)}\Bigg) \nonumber\\
&&\times \int_{0}^{1} dt_1\;t_1^{\frac{1}{2}(-\frac{3}{2}+\lambda) } \int_{0}^{1} du_1\;u_1^{\frac{1}{2}(-1+\lambda )}
 \frac{\left(\frac{1+s_1+\sqrt{s_1^2-2(1-2\widetilde{w}_{2,2}(1-t_1)(1-u_1))s_1+1}}{2}\right)^{-(\frac{1}{4}+\lambda )}}{\sqrt{s_1^2-2(1-2\widetilde{w}_{2,2} (1-t_1)(1-u_1))s_1+1}} \nonumber\\
&&\times \Bigg( \widetilde{w}_{1,2}^{-\frac{\lambda }{2}}\left(  \widetilde{w}_{1,2} \partial _{ \widetilde{w}_{1,2}}\right) \widetilde{w}_{1,2}^{\frac{\lambda }{2}} -\frac{h}{2^4 (1+\rho ^2)}\Bigg) \nonumber\\
&&\times  \sum_{\alpha _0 =0}^{\infty } \frac{(\gamma )_{\alpha _0}}{\alpha _0!} s_0^{\alpha _0} \left\{ c_0 \xi^{\lambda } \sum_{i_0=0}^{\alpha _0} \frac{(-\alpha _0)_{i_0} (\alpha _0+ \frac{1}{4}+\lambda )_{i_0}}{(1+\frac{\lambda }{2})_{i_0}(\frac{3}{4}+ \frac{\lambda }{2})_{i_0}} \widetilde{w}_{1,2} ^{i_0} \right\} \mu^2 \label{eq:33}
\end{eqnarray}
where
\begin{eqnarray}
\widetilde{w}_{1,2} &=& \frac{v_1}{(v_1-1)}\; \frac{\widetilde{w}_{2,2} t_1 u_1}{1- \widetilde{w}_{2,2} v_1 (1-t_1)(1-u_1)}\Bigg|_{\Large v_1=\frac{1-s_1-\sqrt{(1-s_1)^2+4\widetilde{w}_{2,2} (1-t_1)(1-u_1)s_1}}{2\widetilde{w}_{2,2} (1-t_1)(1-u_1)}\normalsize}\nonumber\\
&=& \frac{\widetilde{w}_{2,2} t_1 u_1 \left\{ 1+ (s_1+2\widetilde{w}_{2,2}(1-t_1)(1-u_1) )s_1-(1+s_1)\sqrt{s_1^2-2(1-2\widetilde{w}_{2,2} (1-t_1)(1-u_1))s_1+1}\right\}}{2(1-\widetilde{w}_{2,2}(1-t_1)(1-u_1))^2 s_1}\nonumber
\end{eqnarray}
 Acting the summation operator $\displaystyle{ \sum_{\alpha _0 =0}^{\infty } \frac{(\gamma)_{\alpha _0}}{\alpha _0!} s_0^{\alpha _0} \prod _{n=1}^{\infty } \left\{ \sum_{ \alpha _n = \alpha _{n-1}}^{\infty } s_n^{\alpha _n }\right\}}$ on (\ref{eq:24d}),
\begin{eqnarray}
&&\sum_{\alpha _0 =0}^{\infty } \frac{(\gamma)_{\alpha _0}}{\alpha _0!} s_0^{\alpha _0} \prod _{n=1}^{\infty } \left\{ \sum_{ \alpha _n = \alpha _{n-1}}^{\infty } s_n^{\alpha _n }\right\} y_3(\xi ) \nonumber\\
&&= \prod_{l=4}^{\infty } \frac{1}{(1-s_{l,\infty })} \int_{0}^{1} dt_3\;t_3^{\frac{1}{2}(\frac{1}{2}+\lambda) } \int_{0}^{1} du_3\;u_3^{\frac{1}{2}(1+\lambda )}
 \frac{\left(\frac{1+s_{3,\infty }+\sqrt{s_{3,\infty }^2-2(1-2\eta (1-t_3)(1-u_3))s_{3,\infty }+1}}{2}\right)^{-(\frac{9}{4}+\lambda )}}{\sqrt{s_{3,\infty }^2-2(1-2\eta (1-t_3)(1-u_3))s_{3,\infty }+1}} \nonumber\\
&&\times  \Bigg( \widetilde{w}_{3,3}^{-\frac{1}{2}(2+\lambda )}\left( \widetilde{w}_{3,3} \partial _{ \widetilde{w}_{3,3}}\right) \widetilde{w}_{3,3}^{\frac{1}{2}(2+\lambda )} -\frac{h}{2^4 (1+\rho ^2)}\Bigg) \nonumber\\
&&\times \int_{0}^{1} dt_2\;t_2^{\frac{1}{2}(-\frac{1}{2}+\lambda) } \int_{0}^{1} du_2\;u_2^{\frac{\lambda }{2}}
 \frac{\left(\frac{1+s_2+\sqrt{s_2^2-2(1-2\widetilde{w}_{3,3}(1-t_2)(1-u_2))s_2+1}}{2}\right)^{-(\frac{5}{4}+\lambda )}}{\sqrt{s_2^2-2(1-2\widetilde{w}_{3,3} (1-t_2)(1-u_2))s_2+1}} \nonumber\\
&&\times \Bigg( \widetilde{w}_{2,3}^{-\frac{1}{2}(1+\lambda )}\left(  \widetilde{w}_{2,3} \partial _{ \widetilde{w}_{2,3}}\right) \widetilde{w}_{2,3}^{\frac{1}{2}(1+\lambda )} -\frac{h}{2^4 (1+\rho ^2)}\Bigg) \nonumber\\
&&\times \int_{0}^{1} dt_1\;t_1^{\frac{1}{2}(-\frac{3}{2}+\lambda) } \int_{0}^{1} du_1\;u_1^{\frac{1}{2}(-1+\lambda )}
 \frac{\left(\frac{1+s_1+\sqrt{s_1^2-2(1-2\widetilde{w}_{2,3}(1-t_1)(1-u_1))s_1+1}}{2}\right)^{-(\frac{1}{4}+\lambda )}}{\sqrt{s_1^2-2(1-2\widetilde{w}_{2,3} (1-t_1)(1-u_1))s_1+1}} \nonumber\\
&&\times \Bigg( \widetilde{w}_{1,3}^{-\frac{\lambda }{2}}\left(  \widetilde{w}_{1,3} \partial _{ \widetilde{w}_{1,3}}\right) \widetilde{w}_{1,3}^{\frac{\lambda }{2}} -\frac{h}{2^4 (1+\rho ^2)}\Bigg) \nonumber\\
&&\times \sum_{\alpha _0 =0}^{\infty } \frac{(\gamma )_{\alpha _0}}{\alpha _0!} s_0^{\alpha _0} \left\{ c_0 \xi^{\lambda } \sum_{i_0=0}^{\alpha _0} \frac{(-\alpha _0)_{i_0} (\alpha _0+ \frac{1}{4}+\lambda )_{i_0}}{(1+\frac{\lambda }{2})_{i_0}(\frac{3}{4}+ \frac{\lambda }{2})_{i_0}} \widetilde{w}_{1,3} ^{i_0} \right\} \mu^3 \label{eq:34}
\end{eqnarray}
where
\begin{eqnarray}
\widetilde{w}_{3,3} &=& \frac{v_3}{(v_3-1)}\; \frac{\eta  t_3 u_3}{1- \eta (1-t_3)(1-u_3)v_3}\Bigg|_{\Large v_3=\frac{1-s_{3,\infty }-\sqrt{(1-s_{3,\infty })^2+4\eta (1-t_3)(1-u_3)s_{3,\infty }}}{2\eta (1-t_3)(1-u_3)} \normalsize}\nonumber\\
&=& \frac{\eta t_3 u_3 \left\{ 1+ (s_{3,\infty }+2\eta (1-t_3)(1-u_3) )s_{3,\infty }-(1+s_{3,\infty })\sqrt{s_{3,\infty }^2-2(1-2\eta  (1-t_3)(1-u_3))s_{3,\infty }+1}\right\}}{2(1-\eta (1-t_3)(1-u_3))^2 s_{3,\infty }}\nonumber
\end{eqnarray}
\begin{eqnarray}
\widetilde{w}_{2,3} &=& \frac{v_2}{(v_2-1)}\; \frac{\widetilde{w}_{3,3} t_2 u_2}{1- \widetilde{w}_{3,3} (1-t_2)(1-u_2)v_2 }\Bigg|_{\Large v_2=\frac{1-s_2-\sqrt{(1-s_2)^2+4\widetilde{w}_{3,3} (1-t_2)(1-u_2)s_2}}{2\widetilde{w}_{3,3} (1-t_2)(1-u_2)}\normalsize }\nonumber\\
&=& \frac{\widetilde{w}_{3,3} t_2 u_2 \left\{ 1+ (s_2+2\widetilde{w}_{3,3}(1-t_2)(1-u_2) )s_2-(1+s_2)\sqrt{s_2^2-2(1-2\widetilde{w}_{3,3} (1-t_2)(1-u_2))s_2+1}\right\}}{2(1-\widetilde{w}_{3,3}(1-t_2)(1-u_2))^2 s_2}\nonumber
\end{eqnarray}
\begin{eqnarray}
\widetilde{w}_{1,3} &=& \frac{v_1}{(v_1-1)}\; \frac{\widetilde{w}_{2,3} t_1 u_1}{1- \widetilde{w}_{2,3} (1-t_1)(1-u_1)v_1 }\Bigg|_{\Large v_1=\frac{1-s_1-\sqrt{(1-s_1)^2+4\widetilde{w}_{2,3} (1-t_1)(1-u_1)s_1}}{2\widetilde{w}_{2,3} (1-t_1)(1-u_1)}\normalsize}\nonumber\\
&=& \frac{\widetilde{w}_{2,3} t_1 u_1 \left\{ 1+ (s_1+2\widetilde{w}_{2,3}(1-t_1)(1-u_1) )s_1-(1+s_1)\sqrt{s_1^2-2(1-2\widetilde{w}_{2,3} (1-t_1)(1-u_1))s_1+1}\right\}}{2(1-\widetilde{w}_{2,3}(1-t_1)(1-u_1))^2 s_1}\nonumber
\end{eqnarray}
By repeating this process for all higher terms of integral forms of sub-summation $y_m(\xi )$ terms where $m > 3$, I obtain every  $\displaystyle{ \sum_{\alpha _0 =0}^{\infty } \frac{(\gamma)_{\alpha _0}}{\alpha _0!} s_0^{\alpha _0} \prod _{n=1}^{\infty } \left\{ \sum_{ \alpha _n = \alpha _{n-1}}^{\infty } s_n^{\alpha _n }\right\}}  y_m(\xi )$ terms. 
Substitute (\ref{eq:25}), (\ref{eq:28}), (\ref{eq:33}), (\ref{eq:34}) and including all $\displaystyle{ \sum_{\alpha _0 =0}^{\infty } \frac{(\gamma)_{\alpha _0}}{\alpha _0!} s_0^{\alpha _0} \prod _{n=1}^{\infty } \left\{ \sum_{ \alpha _n = \alpha _{n-1}}^{\infty } s_n^{\alpha _n }\right\}}  y_m(\xi )$ terms where $m > 3$ into (\ref{eq:80}). 
\qed
\end{pot}
\begin{rmk}
The generating function of the first kind of independent solution of Lame polynomial which makes $B_n$ term terminated in Weierstrass's form about $\xi=0 $  as $\alpha =2(2\alpha_j+j) $ or $-2(2\alpha_j+j)-1$ where $j,\alpha _j \in \mathbb{N}_{0}$ is
\begin{eqnarray}
&&\sum_{\alpha _0 =0}^{\infty } \frac{(\frac{3}{4})_{\alpha _0}}{\alpha _0!} s_0^{\alpha _0} \prod _{n=1}^{\infty } \left\{ \sum_{ \alpha _n = \alpha _{n-1}}^{\infty } s_n^{\alpha _n }\right\} LF_{\alpha _j}\Bigg( \rho, h, \alpha = 2(2\alpha_j +j)\; \mbox{or} -2(2\alpha_j +j)-1; \xi = sn^2(z,\rho )\nonumber\\
&&, \mu =(1+\rho ^2) \xi, \eta = -\rho ^2 \xi^2 \Bigg) \nonumber\\
&&=2^{-\frac{3}{4}}\Bigg\{ \prod_{l=1}^{\infty } \frac{1}{(1-s_{l,\infty })} \mathbf{A} \left( s_{0,\infty } ;\eta\right)   \nonumber\\
&&+ \left\{ \prod_{l=2}^{\infty } \frac{1}{(1-s_{l,\infty })} \int_{0}^{1} dt_1\;t_1^{-\frac{3}{4}} \int_{0}^{1} du_1\;u_1^{-\frac{1}{2}} \overleftrightarrow {\mathbf{\Gamma}}_1 \left(s_{1,\infty };t_1,u_1,\eta\right)   \Bigg( \left(  \widetilde{w}_{1,1} \partial _{ \widetilde{w}_{1,1}}\right)^2  -\frac{h}{2^4 (1+\rho ^2)}\Bigg) \mathbf{A} \left(  s_{0} ;\widetilde{w}_{1,1}\right) \right\}\mu \nonumber\\
&&+ \sum_{n=2}^{\infty } \Bigg\{ \prod_{l=n+1}^{\infty } \frac{1}{(1-s_{l,\infty })} \int_{0}^{1} dt_n\;t_n^{\frac{1}{2}(n-\frac{5}{2})} \int_{0}^{1} du_n\;u_n^{\frac{1}{2}(n-2)} \overleftrightarrow {\mathbf{\Gamma}}_n \left(s_{n,\infty };t_n,u_n,\eta \right)\nonumber\\
&&\times \Bigg( \widetilde{w}_{n,n}^{-\frac{1}{2}(n-1)}\left( \widetilde{w}_{n,n} \partial _{ \widetilde{w}_{n,n}}\right)^2 \widetilde{w}_{n,n}^{\frac{1}{2}(n-1)}-\frac{h}{2^4 (1+\rho ^2)}\Bigg) \nonumber\\
&&\times \prod_{k=1}^{n-1} \Bigg\{ \int_{0}^{1} dt_{n-k}\;t_{n-k}^{\frac{1}{2}(n-k-\frac{5}{2})} \int_{0}^{1} du_{n-k} \;u_{n-k}^{\frac{1}{2}(n-k-2)}\overleftrightarrow {\mathbf{\Gamma}}_{n-k} \left(s_{n-k};t_{n-k},u_{n-k},\widetilde{w}_{n-k+1,n} \right)  \nonumber\\
&&\times \left( \widetilde{w}_{n-k,n}^{-\frac{1}{2}(n-k-1)}\left( \widetilde{w}_{n-k,n} \partial _{ \widetilde{w}_{n-k,n}}\right)^2 \widetilde{w}_{n-k,n}^{\frac{1}{2}(n-k-1)}-\frac{h}{2^4 (1+\rho ^2)}\right) \Bigg\} \mathbf{A} \left( s_{0} ;\widetilde{w}_{1,n}\right) \Bigg\} \mu^n \Bigg\}
 \label{eq:9}
\end{eqnarray}
where
\begin{equation}
\begin{cases} 
{ \displaystyle \overleftrightarrow {\mathbf{\Gamma}}_1 \left(s_{1,\infty };t_1,u_1,\eta\right)= \frac{\left( \frac{(1+s_{1,\infty })+\sqrt{s_{1,\infty }^2-2(1-2\eta (1-t_1)(1-u_1))s_{1,\infty }+1}}{2}\right)^{-\frac{1}{4}}}{\sqrt{s_{1,\infty }^2-2(1-2\eta (1-t_1)(1-u_1))s_{1,\infty }+1}} }\cr
{ \displaystyle  \overleftrightarrow {\mathbf{\Gamma}}_n \left(s_{n,\infty };t_n,u_n,\eta \right) = \frac{ \left( \frac{(1+s_{n,\infty })+\sqrt{s_{n,\infty }^2-2(1-2\eta (1-t_n)(1-u_n))s_{n,\infty }+1}}{2}\right)^{-(n-\frac{3}{4})} }{\sqrt{s_{n,\infty }^2-2(1-2\eta (1-t_n)(1-u_n))s_{n,\infty }+1}}}\cr
{ \displaystyle \overleftrightarrow {\mathbf{\Gamma}}_{n-k} \left(s_{n-k};t_{n-k},u_{n-k},\widetilde{w}_{n-k+1,n} \right) = \frac{ \left( \frac{(1+s_{n-k})+\sqrt{s_{n-k}^2-2(1-2\widetilde{w}_{n-k+1,n} (1-t_{n-k})(1-u_{n-k}))s_{n-k}+1}}{2}\right)^{-(n-k-\frac{3}{4})} }{\sqrt{s_{n-k}^2-2(1-2\widetilde{w}_{n-k+1,n} (1-t_{n-k})(1-u_{n-k}))s_{n-k}+1}}}
\end{cases}\nonumber 
\end{equation}
and
\begin{equation}
\begin{cases} 
{ \displaystyle \mathbf{A} \left( s_{0,\infty } ;\eta\right)= \frac{\left(1- s_{0,\infty }+\sqrt{s_{0,\infty }^2-2(1-2\eta )s_{0,\infty }+1}\right)^{\frac{1}{4}} \left(1+s_{0,\infty }+\sqrt{s_{0,\infty }^2-2(1-2\eta )s_{0,\infty }+1}\right)^{\frac{1}{2}}}{\sqrt{s_{0,\infty }^2-2(1-2\eta )s_{0,\infty }+1}}}\cr
{ \displaystyle  \mathbf{A} \left( s_{0} ;\widetilde{w}_{1,1}\right) = \frac{\left(1- s_0+\sqrt{s_0^2-2(1-2\widetilde{w}_{1,1})s_0+1}\right)^{\frac{1}{4}} \left(1+s_0+\sqrt{s_0^2-2(1-2\widetilde{w}_{1,1} )s_0+1}\right)^{\frac{1}{2}}}{\sqrt{s_0^2-2(1-2\widetilde{w}_{1,1})s_0+1}}} \cr
{ \displaystyle \mathbf{A} \left( s_{0} ;\widetilde{w}_{1,n}\right) = \frac{\left(1- s_0+\sqrt{s_0^2-2(1-2\widetilde{w}_{1,n})s_0+1}\right)^{\frac{1}{4}} \left(1+s_0+\sqrt{s_0^2-2(1-2\widetilde{w}_{1,n} )s_0+1}\right)^{\frac{1}{2}}}{\sqrt{s_0^2-2(1-2\widetilde{w}_{1,n})s_0+1}}}
\end{cases}\nonumber 
\end{equation}
\end{rmk}
\begin{pf}
Replace $\gamma $, A, w and x  by $3/4$, $1/4$, $s_{0,\infty }$ and $\eta $ in (\ref{eq:3}). 
\begin{eqnarray}
&&\sum_{\alpha _0=0}^{\infty }\frac{(\frac{3}{4})_{\alpha _0}}{\alpha _0!} s_{0,\infty }^{\alpha _0} \;_2F_1\left(-\alpha _0, \alpha _0+\frac{1}{4}; \frac{3}{4}; \eta \right) \label{eq:35}\\
&&= 2^{-\frac{3}{4}}\frac{\left(1- s_{0,\infty }+\sqrt{s_{0,\infty }^2-2(1-2\eta )s_{0,\infty }+1}\right)^{\frac{1}{4}} \left(1+s_{0,\infty }+\sqrt{s_{0,\infty }^2-2(1-2\eta )s_{0,\infty }+1}\right)^{\frac{1}{2}}}{\sqrt{s_{0,\infty }^2-2(1-2\eta )s_{0,\infty }+1}} \nonumber
\end{eqnarray} 
Replace $\gamma $, A, w and x  by $3/4$, $1/4$, $s_0$ and $\widetilde{w}_{1,1}$ in (\ref{eq:3}). 
\begin{eqnarray}
&&\sum_{\alpha _0=0}^{\infty }\frac{(\frac{3}{4})_{\alpha _0}}{\alpha _0!} s_0^{\alpha _0} \;_2F_1\left(-\alpha _0, \alpha _0+\frac{1}{4}; \frac{3}{4}; \widetilde{w}_{1,1} \right) \label{eq:36}\\
&&= 2^{-\frac{3}{4}}\frac{\left(1- s_0+\sqrt{s_0^2-2(1-2\widetilde{w}_{1,1})s_0+1}\right)^{\frac{1}{4}} \left(1+s_0+\sqrt{s_0^2-2(1-2\widetilde{w}_{1,1} )s_0+1}\right)^{\frac{1}{2}}}{\sqrt{s_0^2-2(1-2\widetilde{w}_{1,1})s_0+1}} \nonumber
\end{eqnarray} 
Replace $\gamma $, A, w and x  by $3/4$, $1/4$, $s_0$ and $\widetilde{w}_{1,n}$ in (\ref{eq:3}). 
\begin{eqnarray}
&&\sum_{\alpha _0=0}^{\infty }\frac{(\frac{3}{4})_{\alpha _0}}{\alpha _0!} s_0^{\alpha _0} \;_2F_1\left(-\alpha _0, \alpha _0+\frac{1}{4}; \frac{3}{4}; \widetilde{w}_{1,n} \right) \label{eq:37}\\
&&= 2^{-\frac{3}{4}}\frac{\left(1- s_0+\sqrt{s_0^2-2(1-2\widetilde{w}_{1,n})s_0+1}\right)^{\frac{1}{4}} \left(1+s_0+\sqrt{s_0^2-2(1-2\widetilde{w}_{1,n} )s_0+1}\right)^{\frac{1}{2}}}{\sqrt{s_0^2-2(1-2\widetilde{w}_{1,n})s_0+1}} \nonumber
\end{eqnarray} 
Put $c_0$= 1, $\lambda $=0 and $\gamma=3/4$ in (\ref{eq:8}). Substitute (\ref{eq:35}), (\ref{eq:36}) and (\ref{eq:37}) into the new (\ref{eq:8}).
\qed
\end{pf}
\begin{rmk}
The generating function of the second kind of independent solution of Lame polynomial which makes $B_n$ term terminated in Weierstrass's form about $\xi =0 $ as $\alpha =2(2\alpha_j+j)+1 $ or $-2(2\alpha_j+j+1)$ where $j,\alpha _j \in \mathbb{N}_{0}$ is
\begin{eqnarray}
&&\sum_{\alpha _0 =0}^{\infty } \frac{(\frac{5}{4})_{\alpha _0}}{\alpha _0!} s_0^{\alpha _0} \prod _{n=1}^{\infty } \left\{ \sum_{ \alpha _n = \alpha _{n-1}}^{\infty } s_n^{\alpha _n }\right\} LS_{\alpha _j}\Bigg( \rho, h, \alpha =  2(2\alpha_j +j)+1\; \mbox{or} -2(2\alpha_j +j+1); \xi = sn^2(z,\rho )\nonumber\\
&&, \mu = (1+\rho ^2) \xi, \eta = -\rho ^2 \xi^2 \Bigg) \nonumber\\
&=& \left(\frac{\xi^2}{2}\right)^{\frac{1}{4}} \Bigg\{ \prod_{l=1}^{\infty } \frac{1}{(1-s_{l,\infty })} \mathbf{B} \left( s_{0,\infty } ;\eta\right)\nonumber\\
&+& \left\{ \prod_{l=2}^{\infty } \frac{1}{(1-s_{l,\infty })} \int_{0}^{1} dt_1\;t_1^{-\frac{1}{2}} \int_{0}^{1} du_1\;u_1^{-\frac{1}{4}}  \overleftrightarrow {\mathbf{\Psi }}_1 \left(s_{1,\infty };t_1,u_1,\eta\right) \Bigg(  \widetilde{w}_{1,1}^{-\frac{1}{4}}\left(  \widetilde{w}_{1,1} \partial _{ \widetilde{w}_{1,1}}\right)^2 \widetilde{w}_{1,1}^{\frac{1}{4}} -\frac{h}{2^4 (1+\rho ^2)}\Bigg) \mathbf{B} \left( s_{0} ;\widetilde{w}_{1,1}\right)\right\} \mu \nonumber\\
&+& \sum_{n=2}^{\infty } \Bigg\{ \prod_{l=n+1}^{\infty } \frac{1}{(1-s_{l,\infty })} \int_{0}^{1} dt_n\;t_n^{\frac{1}{2}(n-2)} \int_{0}^{1} du_n\;u_n^{\frac{1}{2}(n-\frac{3}{2} )} \overleftrightarrow {\mathbf{\Psi }}_n \left(s_{n,\infty };t_n,u_n,\eta \right)\nonumber\\
&\times&  \Bigg(  \widetilde{w}_{n,n}^{-\frac{1}{2}(n-\frac{1}{2})}\left( \widetilde{w}_{n,n} \partial _{ \widetilde{w}_{n,n}}\right)^2 \widetilde{w}_{n,n}^{\frac{1}{2}(n-\frac{1}{2})}-\frac{h}{2^4 (1+\rho ^2)}\Bigg) \nonumber\\
&\times& \prod_{k=1}^{n-1} \Bigg\{ \int_{0}^{1} dt_{n-k}\;t_{n-k}^{\frac{1}{2}(n-k-2)} \int_{0}^{1} du_{n-k} \;u_{n-k}^{\frac{1}{2}(n-k-\frac{3}{2})}\overleftrightarrow {\mathbf{\Psi }}_{n-k} \left(s_{n-k};t_{n-k},u_{n-k},\widetilde{w}_{n-k+1,n} \right) \nonumber\\
&\times& \Bigg( \widetilde{w}_{n-k,n}^{-\frac{1}{2}(n-k-\frac{1}{2})}\left( \widetilde{w}_{n-k,n} \partial _{ \widetilde{w}_{n-k,n}}\right)^2 \widetilde{w}_{n-k,n}^{\frac{1}{2}(n-k-\frac{1}{2})}-\frac{h}{2^4 (1+\rho ^2)}\Bigg) \Bigg\} \mathbf{B} \left(  s_{0} ;\widetilde{w}_{1,n}\right)\Bigg\} \mu^n  \Bigg\} \hspace{1cm}\label{eq:10}
\end{eqnarray}
where
\begin{equation}
\begin{cases} 
{ \displaystyle \overleftrightarrow {\mathbf{\Psi }}_1 \left(s_{1,\infty };t_1,u_1,\eta\right)=  \frac{\left( \frac{(1+s_{1,\infty })+\sqrt{s_{1,\infty }^2-2(1-2\eta (1-t_1)(1-u_1))s_{1,\infty }+1}}{2}\right)^{-\frac{3}{4}}}{\sqrt{s_{1,\infty }^2-2(1-2\eta (1-t_1)(1-u_1))s_{1,\infty }+1}}}\cr
{ \displaystyle  \overleftrightarrow {\mathbf{\Psi }}_n \left(s_{n,\infty };t_n,u_n,\eta \right) =  \frac{ \left( \frac{(1+s_{n,\infty })+\sqrt{s_{n,\infty }^2-2(1-2\eta (1-t_n)(1-u_n))s_{n,\infty }+1}}{2}\right)^{-(n-\frac{1}{4})} }{\sqrt{s_{n,\infty }^2-2(1-2\eta (1-t_n)(1-u_n))s_{n,\infty }+1}}}\cr
{ \displaystyle \overleftrightarrow {\mathbf{\Psi }}_{n-k} \left(s_{n-k};t_{n-k},u_{n-k},\widetilde{w}_{n-k+1,n} \right) = \frac{ \left( \frac{(1+s_{n-k})+\sqrt{s_{n-k}^2-2(1-2\widetilde{w}_{n-k+1,n} (1-t_{n-k})(1-u_{n-k}))s_{n-k}+1}}{2}\right)^{-(n-k-\frac{1}{4})} }{\sqrt{s_{n-k}^2-2(1-2\widetilde{w}_{n-k+1,n} (1-t_{n-k})(1-u_{n-k}))s_{n-k}+1}} }
\end{cases}\nonumber 
\end{equation}
and
\begin{equation}
\begin{cases} 
{ \displaystyle \mathbf{B} \left( s_{0,\infty } ;\eta\right)= \frac{\left(1- s_{0,\infty }+\sqrt{s_{0,\infty }^2-2(1-2\eta )s_{0,\infty }+1}\right)^{-\frac{1}{4}} \left(1+s_{0,\infty }+\sqrt{s_{0,\infty }^2-2(1-2\eta )s_{0,\infty }+1}\right)^{\frac{1}{2}}}{\sqrt{s_{0,\infty }^2-2(1-2\eta )s_{0,\infty }+1}}}\cr
{ \displaystyle  \mathbf{B} \left( s_{0} ;\widetilde{w}_{1,1}\right) = \frac{\left(1- s_0+\sqrt{s_0^2-2(1-2\widetilde{w}_{1,1})s_0+1}\right)^{-\frac{1}{4}} \left(1+s_0+\sqrt{s_0^2-2(1-2\widetilde{w}_{1,1} )s_0+1}\right)^{\frac{1}{2}}}{\sqrt{s_0^2-2(1-2\widetilde{w}_{1,1})s_0+1}}} \cr
{ \displaystyle \mathbf{B} \left( s_{0} ;\widetilde{w}_{1,n}\right) = \frac{\left(1- s_0+\sqrt{s_0^2-2(1-2\widetilde{w}_{1,n})s_0+1}\right)^{-\frac{1}{4}} \left(1+s_0+\sqrt{s_0^2-2(1-2\widetilde{w}_{1,n} )s_0+1}\right)^{\frac{1}{2}}}{\sqrt{s_0^2-2(1-2\widetilde{w}_{1,n})s_0+1}}}
\end{cases}\nonumber 
\end{equation}
\end{rmk}
\begin{pf}
Replace $\gamma $, A, w and x  by $5/4$, $3/4$, $s_{0,\infty }$ and $\eta $ in (\ref{eq:3}). 
\begin{eqnarray}
&&\sum_{\alpha _0=0}^{\infty }\frac{(\frac{5}{4})_{\alpha _0}}{\alpha _0!} s_{0,\infty }^{\alpha _0} \;_2F_1\left(-\alpha _0, \alpha _0+\frac{3}{4}; \frac{5}{4}; \eta \right) \label{eq:38}\\
&&= 2^{-\frac{1}{4}}\frac{\left(1- s_{0,\infty }+\sqrt{s_{0,\infty }^2-2(1-2\eta )s_{0,\infty }+1}\right)^{-\frac{1}{4}} \left(1+s_{0,\infty }+\sqrt{s_{0,\infty }^2-2(1-2\eta )s_{0,\infty }+1}\right)^{\frac{1}{2}}}{\sqrt{s_{0,\infty }^2-2(1-2\eta )s_{0,\infty }+1}} \nonumber
\end{eqnarray} 
Replace $\gamma $, A, w and x  by $5/4$, $3/4$, $s_0$ and $\widetilde{w}_{1,1}$ in (\ref{eq:3}). 
\begin{eqnarray}
&&\sum_{\alpha _0=0}^{\infty }\frac{(\frac{5}{4})_{\alpha _0}}{\alpha _0!} s_0^{\alpha _0} \;_2F_1\left(-\alpha _0, \alpha _0+\frac{3}{4}; \frac{5}{4}; \widetilde{w}_{1,1} \right) \label{eq:39}\\
&&= 2^{-\frac{1}{4}}\frac{\left(1- s_0+\sqrt{s_0^2-2(1-2\widetilde{w}_{1,1})s_0+1}\right)^{-\frac{1}{4}} \left(1+s_0+\sqrt{s_0^2-2(1-2\widetilde{w}_{1,1} )s_0+1}\right)^{\frac{1}{2}}}{\sqrt{s_0^2-2(1-2\widetilde{w}_{1,1})s_0+1}} \nonumber
\end{eqnarray} 
Replace $\gamma $, A, w and x  by $5/4$, $3/4$, $s_0$ and $\widetilde{w}_{1,n}$ in (\ref{eq:3}). 
\begin{eqnarray}
&&\sum_{\alpha _0=0}^{\infty }\frac{(\frac{5}{4})_{\alpha _0}}{\alpha _0!} s_0^{\alpha _0} \;_2F_1\left(-\alpha _0, \alpha _0+\frac{3}{4}; \frac{5}{4}; \widetilde{w}_{1,n} \right) \label{eq:40}\\
&&= 2^{-\frac{1}{4}}\frac{\left(1- s_0+\sqrt{s_0^2-2(1-2\widetilde{w}_{1,n})s_0+1}\right)^{-\frac{1}{4}} \left(1+s_0+\sqrt{s_0^2-2(1-2\widetilde{w}_{1,n} )s_0+1}\right)^{\frac{1}{2}}}{\sqrt{s_0^2-2(1-2\widetilde{w}_{1,n})s_0+1}} \nonumber
\end{eqnarray} 
Put $c_0= 1$, $\lambda =1/2$ and $\gamma=5/4$ in (\ref{eq:8}). Substitute (\ref{eq:38}), (\ref{eq:39}) and (\ref{eq:40}) into the new (\ref{eq:8}).
\qed
\end{pf}

\section{Summary} 
In my previous two papers: (1) I show the power series expansion in closed forms and integral form of Lame function in the algebraic form (infinite series and polynomial which makes $B_n$ term terminated) and its asymptotic behaviors\cite{Chou2012f}. (2) By applying three term recurrence formula \cite{chou2012b}, I construct the power series expansion and integral form of Lame function in the the Weierstrass's form (infinite series and polynomial which makes $B_n$ term terminated) and its asymptotic behaviors\cite{Chou2012g}. In this paper I derive generating function of Lame polynomial which makes $B_n$ term terminated in Weierstrass's form including all higher terms of $A_n$'s. We can apply power series expansion in closed forms, integral form and generating function of Lame polynomial in many cases of modern physics\cite{Brac2001,Qian2003,Maie2001,Capu2000,Kant2001,Dobn1998,Iach2000}.\footnote{ In mathematical definition, Lame polynomial or Lame spectral polynomial is type 3 polynomial where $A_n$ and $B_n$ terms terminated: for any non-negative integer value of $\alpha $ there will be $2\alpha  + 1$ values of $h$ for which the solution $y(\xi )$ reduces to a polynomial. In this paper I construct the generating function of Lame polynomial of type 1: I treat the spectral parameter $h$ as a free variable. In my next papers I will work on the generating functions of Lame polynomial of type 2.}

I construct the power series expansion in closed forms and analytic integral solution of linear ordinary differential equations that have three term recursion relations by using the three term recurrence formula\cite{chou2012b}: such as Heun\cite{chou2012c,Chou2012d}, Lame functions\cite{Chou2012f,Chou2012g}, Grand Confluent Hypergeometric (GCH)\cite{Chou2012i,Chou2012j} and Mathieu\cite{Chou2012e}.
As we see all solutions of power series expansions in GCH, Mathieu, Heun and Lame functions by using the three-term recurrence formula \cite{chou2012b}, denominators and numerators in all $B_n$ terms arise with Pochhammer symbol: the meaning of this is that the analytic solutions of any ordinary differential equations with three recursive coefficients can be described as Hypergoemetric function in a strict mathematical way. 

We can express representations in closed form integrals in an easy way since we have power series expansions with Pochhammer symbols in numerators and denominators. We can transform any special functions, having three term recursive relation, into all other well-known special functions with two recursive coefficients because a $_2F_1$ function recurs in each of sub-integral forms of them. It means all analytic solutions in the three-term recurrence can be described as Hypergoemetric function: understanding the connection between other special functions is important in the mathematical and physical points of views as we all know.

Since analytic integral form of ordinary differential equations with three recursive coefficients are derived from power series expansion in closed forms, we might be possible to construct generating function of all these polynomials. The generating function is really helpful in order to derive orthogonal relations, recursion relations and expectation values of any physical quantities as we all recognize; i.e. the normalized wave function of hydrogen-like atoms and expectation values of its physical quantities such as position and momentum.

\section{Series ``Special functions and three term recurrence formula (3TRF)''} 

This paper is 8th out of 10.
\vspace{3mm}

1. ``Approximative solution of the spin free Hamiltonian involving only scalar potential for the $q-\bar{q}$ system'' \cite{chou2012a} - In order to solve the spin-free Hamiltonian with light quark masses we are led to develop a totally new kind of special function theory in mathematics that generalize all existing theories of confluent hypergeometric types. We call it the Grand Confluent Hypergeometric Function. Our new solution produces previously unknown extra hidden quantum numbers relevant for description of supersymmetry and for generating new mass formulas.
\vspace{3mm}

2. ``Generalization of the three-term recurrence formula and its applications'' \cite{chou2012b} - Generalize three term recurrence formula in linear differential equation.  Obtain the exact solution of the three term recurrence for polynomials and infinite series.
\vspace{3mm}

3. ``The analytic solution for the power series expansion of Heun function'' \cite{chou2012c} -  Apply three term recurrence formula to the power series expansion in closed forms of Heun function (infinite series and polynomials) including all higher terms of $A_n$'s.
\vspace{3mm}

4. ``Asymptotic behavior of Heun function and its integral formalism'', \cite{Chou2012d} - Apply three term recurrence formula, derive the integral formalism, and analyze the asymptotic behavior of Heun function (including all higher terms of $A_n$'s). 
\vspace{3mm}

5. ``The power series expansion of Mathieu function and its integral formalism'', \cite{Chou2012e} - Apply three term recurrence formula, analyze the power series expansion of Mathieu function and its integral forms.  
\vspace{3mm}

6. ``Lame equation in the algebraic form'' \cite{Chou2012f} - Applying three term recurrence formula, analyze the power series expansion of Lame function in the algebraic form and its integral forms.
\vspace{3mm}

7. ``Power series and integral forms of Lame equation in   Weierstrass's form and its asymptotic behaviors'' \cite{Chou2012g} - Applying three term recurrence formula, derive the power series expansion of Lame function in   Weierstrass's form and its integral forms. 
\vspace{3mm}

8. ``The generating functions of Lame equation in   Weierstrass's form'' \cite{Chou2012h} - Derive the generating functions of Lame function in   Weierstrass's form (including all higher terms of $A_n$'s).  Apply integral forms of Lame functions in   Weierstrass's form.
\vspace{3mm}

9. ``Analytic solution for grand confluent hypergeometric function'' \cite{Chou2012i} - Apply three term recurrence formula, and formulate the exact analytic solution of grand confluent hypergeometric function (including all higher terms of $A_n$'s). Replacing $\mu $ and $\varepsilon \omega $ by 1 and $-q$, transforms the grand confluent hypergeometric function into Biconfluent Heun function.
\vspace{3mm}

10. ``The integral formalism and the generating function of grand confluent hypergeometric function'' \cite{Chou2012j} - Apply three term recurrence formula, and construct an integral formalism and a generating function of grand confluent hypergeometric function (including all higher terms of $A_n$'s). 
\section*{Acknowledgment}
I thank Bogdan Nicolescu. The discussions I had with him on number theory was of great joy. 
\vspace{3mm}

\bibliographystyle{model1a-num-names}
\bibliography{<your-bib-database>}
 
\end{document}